\documentclass[12pt]{article}

\usepackage{textcomp}
\usepackage{scicite}
\usepackage{gensymb}
\usepackage{graphicx}
\usepackage[utf8]{inputenc}
\usepackage{lmodern}
\usepackage{multibib}
\usepackage{caption}
\usepackage{gensymb}
\usepackage{ccaption}
\usepackage[rightcaption]{sidecap}
\sidecaptionvpos{figure}{t}
\usepackage{amsmath}

\usepackage{times}

\newenvironment{sciabstract}{%
\begin{quote} \bf}
{\end{quote}}

\newcounter{lastnote}

\title{Hexatic phase in covalent two-dimensional \\ silver iodide}

\author
{Thuy An Bui $^{1\dag}$, David Lamprecht $^{1,2\dag}$, Jacob Madsen $^{1}$, Marcin Kurpas,$^{3}$ \\ Peter Kotrusz,$^{4,7}$ Alexander Markevich,$^{1}$ Clemens Mangler,$^{1}$ \\  Jani Kotakoski, $^{1}$ Lado Filipovic,$^{2}$ Jannik C. Meyer,$^{5}$ \\ Timothy J. Pennycook,$^{6}$ Viera Sk\'akalov\'a,$^{1,4,7}$ Kimmo Mustonen$^{1\ast}$\\
\\
\normalsize{$^{1}$University of Vienna, Faculty of Physics,}
\normalsize{1090 Vienna, Austria}\\
\normalsize{$^{2}$Institute for Microelectronics, TU Vienna, }
\normalsize{1040 Vienna, Austria}\\
\normalsize{$^{3}$Institute of Physics, University of Silesia in Katowice,}
\normalsize{41-500 Chorz\'{o}w, Poland}\\
\normalsize{$^{4}$Danubia NanoTech s.r.o., 84104 Bratislava, Slovakia}\\
\normalsize{$^{5}$Eberhard Karls University of Tuebingen, Institute of Applied Physics,}\\
\normalsize{72076 Tuebingen, Germany}\\
\normalsize{$^{6}$University of Antwerp, EMAT, 2020 Antwerp, Belgium}\\
\normalsize{$^{7}$Institute of Electrical Engineering of SAS, 84104 Bratislava, Slovakia}\\
\normalsize{$^{\dag}$Contributed equally to this work} \\
\\
\normalsize{$^\ast$To whom correspondence should be addressed; E-mail: kimmo.mustonen@univie.ac.at}
}

\date{}

\begin{document} 


\baselineskip24pt


\maketitle


\begin{sciabstract}
According to the Kosterlitz-Thouless-Halperin-Nelson-Young (KTHNY) theory, the transition from a solid to liquid in two dimensions (2D) proceeds through an orientationally ordered liquid-like hexatic phase. However, alternative mixed-melting scenarios, in which melting proceeds through the hexatic phase with both continuous and discontinuous transitions, have also been observed in some 2D systems. Here, we imaged silver iodide (AgI) embedded in multilayer graphene using time and temperature resolved in-situ atomic-resolution scanning transmission microscopy (STEM) and nanobeam electron diffraction. We observe the hexatic phase and provide evidence supporting a mixed melting scenario.
\end{sciabstract}

First-order phase transitions in three-dimensional materials are characterized by discontinuous (first-order) changes in thermodynamic quantities such as entropy. A common example is ice melting into water accompanied by the immediate loss of its crystal symmetries. However, in two-dimensional (2D) systems, additional melting scenarios have been proposed. For example, the Kosterlitz–Thouless (KT) transition predicted by the Kosterlitz-Thouless-Halperin-Nelson-Young (KTHNY) theory~\cite{kosterlitz_long_1972, kosterlitz_ordering_1973, halperin_theory_1978, nelson_dislocation-mediated_1979, young_melting_1979, guillamon_direct_2009}, describes a continuous two-step process occurring via an orientationally ordered, liquid-like intermediate phase known as the hexatic phase. Various mixed-melting scenarios ~\cite{fisher_defects_1979, chui_grain-boundary_1982, chui_grain-boundary_1983} propose a continuous transition to the hexatic followed by a first-order transition to a liquid phase~\cite{bernard_two-step_2011}. In addition, direct first-order transitions without an intermediate phase have also been observed in some 2D systems~\cite{armstrong_isothermal-expansion_1989}.

The appearance of the hexatic phase has been experimentally verified in a wide variety of model systems governed by distinct interparticle forces. These include systems dominated by Coulomb repulsion, such as electrons on liquid helium~\cite{guo_evidence_1983}, charged colloids in water~\cite{zheng_melting_2006,murray_experimental_1987}, and dusty plasmas~\cite{vasilieva_laser-induced_2021}; systems dominated by van der Waals interactions, including liquid-crystal and layered molecular films~\cite{geer_liquid-hexatic_1992,viswanathan_liquid_1995,chou_multiple-step_1998,heiney_freezing_1982}; and those governed by dipolar or elastic forces, such as magnetic colloids~\cite{zahn_two-stage_1999}, magnetic skyrmion lattices~\cite{huang_melting_2020}, and biological cell monolayers~\cite{pasupalak_hexatic_2020}. Nevertheless, even in these well-controlled systems, simulations~\cite{li_attraction_2020} and experiments~\cite{thorneywork_two-dimensional_2017,rosenbaum_experimental_1983} show that melting can proceed via different routes—either a grain-boundary-mediated process or the KTHNY mechanism—depending sensitively on particle density and interaction range.
The KTHNY melting of a system with a more complex directional interaction structure has been experimentally observed in the case of charge density waves, both in real space using scanning probe techniques~\cite{dai_solid-hexatic-liquid_1992} and in reciprocal space using ultrafast transmission electron microscopy (TEM)~\cite{domrose_light-induced_2023}. The closest analogue to a covalently bonded system has been provided by simulations of 2D water ice under high pressure \cite{kapil_first-principles_2022}. Although this study revealed an intriguing mixed melting pathway with subsequent first-order and KT transitions, the hydrogen bonds of ice are far weaker than covalent bonds and are pointed exclusively in the in-plane direction. In covalent crystals, the force interactions are not only highly directional but also much stronger than van der Waals interactions or hydrogen bonds, and so far, determining their melting mechanism has remained out of reach for both simulations and experiments. 

Here, we study the dynamic melting of covalently bound hexagonal monolayers of AgI encapsulated by graphene\cite{haastrup_computational_2018, mustonen_toward_2022}. Graphene encapsulation stabilizes 2D AgI and prevents its transformation into 3D phases \cite{hull_pressure-induced_1999}. The non-commensurability of the graphene and AgI lattices eliminates periodic interactions and allows AgI to melt without appreciable orientational constraints in our time- and temperature-resolved in situ scanning transmission electron microscopy (STEM) experiments. 
Convolutional neural network (CNN) analysis of the translational and orientational correlations in our large dataset of experimental images confirmed the existence of the hexatic phase in a covalent material. They also provide evidence for evaluating the proposed melting scenarios for covalent materials, and point to a mixed melting scenario as the most likely one.

\clearpage

\section*{Melting of 2D crystalline silver iodide}

\begin{figure}[!b]
    \centering
    \includegraphics[width=\textwidth]{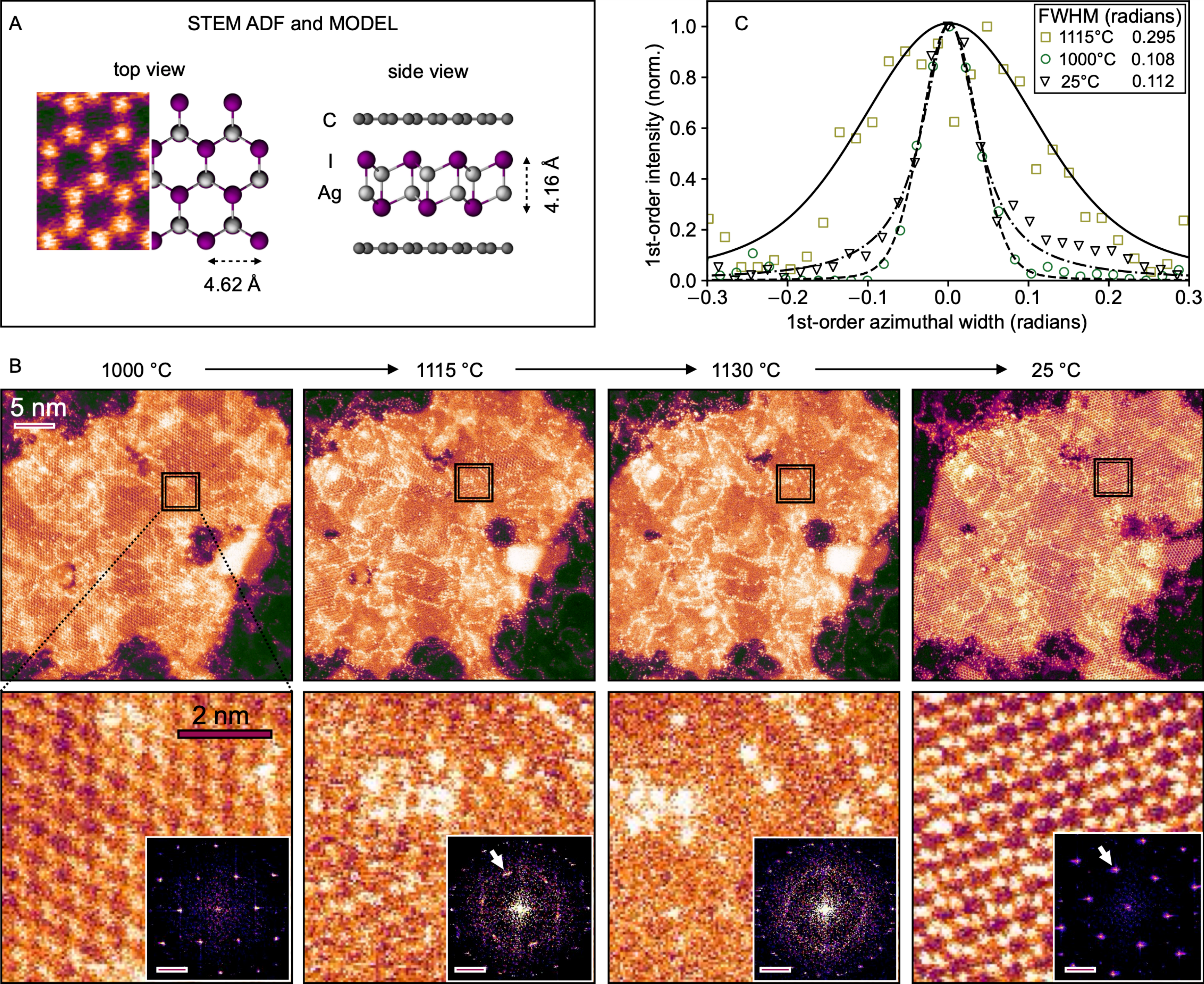}
    \caption{\textbf{Annular dark-field imaging of the melting of 2D crystalline silver iodide.}~\textbf{(A)} Density functional theory (DFT) model of 2D AgI shown beside an atomically resolved ADF image of the room-temperature structure. \textbf{(B)} Development of the AgI structure during heating and cooling: crystalline state at 1000~\degree C, phase one of the 2D melting process at 1115~\degree C with interspersed disordered crystalline and molten areas, and the melting phase two at 1130~\degree C. After cooling down to 25~\degree C, the AgI crystal appears again but is rotated by 12\degree. The bright spots visible on and around the AgI crystal are iodine atoms stuck at the edges of graphitized contamination (see Figures \ref{fig:I_adsorption} and \ref{fig:iodine_STEM}). The FTs in the bottom row are from 100 rapidly acquired STEM images (see Methods). The contrast of the FTs has been optimized to emphasize the AgI peaks. Additional temperature datapoints and larger field of views can be found in figures \ref{fig:2D_melting_extras_1}--\ref{fig:2D_melting_extras_4}. \textbf{(C)} Temperature dependence of the azimuthal width of the first-order AgI peaks marked with arrows in the Fourier transforms (FTs) of the ADF images in panel B. }
    \label{fig:2D_melting}
\end{figure}

The experimental (STEM) and theoretical (density functional theory, DFT) structures of monolayer AgI are shown in
Figure~\ref{fig:2D_melting}A (also Figures \ref{fig: EELS} and \ref{fig:tilt}). The evolution of a single encapsulated crystal as it is heated until it melts, and is then cooled, is shown in~\ref{fig:2D_melting}B . The fully crystalline material transitioned upon heating into a dynamic phase, with regions of disorder appearing and disappearing over time. Then, as the material was heated further, the disorder fully took over. We show below that the first stage manifested as the solid-hexatic phase boundary was crossed, whereas the second stage signified a fully liquid state of the covalent 2D material. These distinct stages of melting are illustrated by Supplementary Movies S1–S3, which also display the CNN-detected positions of atoms (see Methods).

%

The temperatures of the transitions depended on the AgI crystal size (figure~\ref{fig:temperature}). Being embedded in graphene, the largest 2D crystals (\textgreater100 nm in size) melt at approximately 1200~\degree C, which was substantially higher than the 660~\degree C reported for bulk AgI~\cite{hull_pressure-induced_1999}. Nevertheless, crystals of all sizes and shapes consistently proceeded through the two distinct stages of melting, and then fully reverted to 2D AgI after the temperature is lowered sufficiently. Here we primarily focused on the larger AgI crystals, typically 40 to 80 nm in diameter, where edge effects were less prominent. An example of a small AgI crystal undergoing the phase transition is shown in Figure \ref{fig: AgI_small}.

The changes occurring in stages one and two were visible directly in our real-space images and in their Fourier transforms (FT, Methods). The dynamic crystalline phase typically emerged at temperatures between 1000~\degree and 1125~\degree C, and then gradually transformed into a fully molten structure between 1050~\degree and 1200~\degree C, depending on the crystal size. The AgI crystal (Figure~\ref{fig:2D_melting}B at 1115~\degree C) exhibited regions with disordered hexagonal lattice still present, interspersed with areas that appeared to lack obvious crystallinity. The encapsulating graphene was not apparent in the annular dark field images (ADF, Methods) because of its low atomic number~\cite{krivanek_atom-by-atom_2010}, but is distinguished in the FTs, especially at 1130~\degree C, by a set of spots on an outer circle revealing six layers of graphene rotated at different angles in this sample. The graphene layers were unaffected by these temperatures that were well below the melting point of graphene~\cite{zakharchenko_melting_2011}, as seen through the constant sharpness of its characteristic spots in the FT of the images.

The partial loss of orientational order, qualitatively visible in the ADF images during stage one, became evident in the FTs, where the 2D AgI crystal spots broadened in the azimuthal direction as shown in Figure \ref{fig:2D_melting}C. This azimuthal broadening of the first-order peaks was consistent with the presence of a hexatic-like phase slightly below the final melting temperature~\cite{zaluzhnyy_spatially_2015, rosenbaum_experimental_1983}. In this stage of the melting, the material is highly dynamic with the regions of disorder appearing and disappearing over time as captured in the fluctuations in the ADF images shown in Supplementary Movie S2.

At 1130~\degree C the AgI crystallinity had completely vanished (figure~\ref{fig:2D_melting}B and Video S3), indicating that the AgI had become a 2D liquid. Weak isotropic rings of disordered AgI were still visible in the FT, in addition to the higher frequency spots of the unaffected graphene layers. However, the fluctuations seen in the lower temperature dynamic crystalline phase now vanished, leaving only constant isotropic rings in the FTs. We attributed this complete loss of orientational order and presence of only a very short-range translational order, qualitatively different from the one described before, to the liquid phase. Upon cooling to 25~\degree C, the 2D AgI crystal reappeared, rotated by 12\degree~from its original orientation, with sharp FT spots indicating a fully crystalline phase. These rotations were often observed, but they appeared to be random even within the same sample area. This observation was in good agreement with computations that revealed only very weak interaction between AgI and graphene (see Figure~\ref{fig:en_map}), thus ruling out long-range periodic effects that could lead to artifacts in our data.

\begin{figure}[!b]
    \centering  \includegraphics[width=\textwidth]{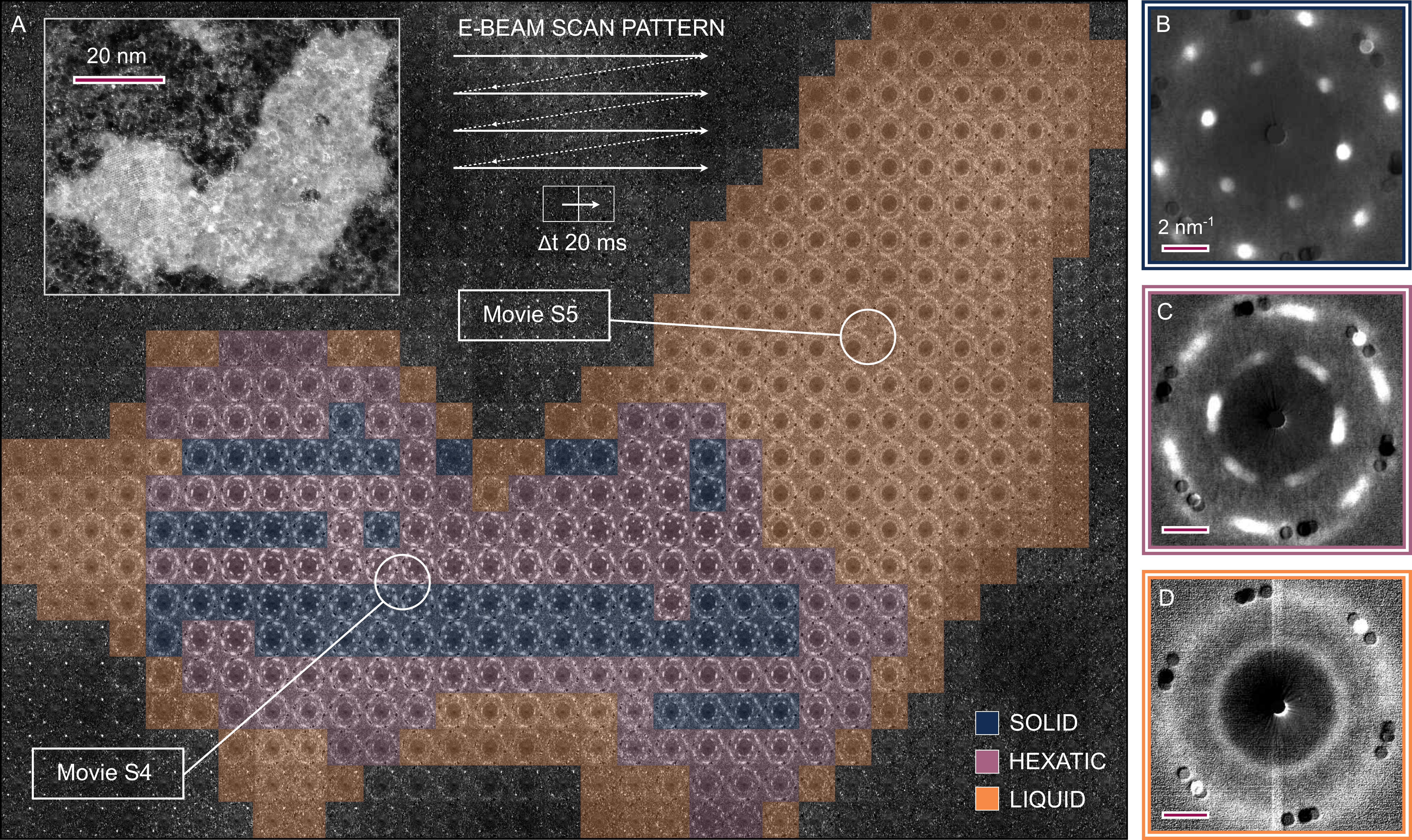}
    \caption{\textbf{Phase analysis based on nanobeam electron diffraction (NBED).}~\textbf{(A)} NBED map of an AgI crystal with its right side in a liquid state and its left side in a dynamic state fluctuating between solid and hexatic phases. Supplementary Movies S4 and S5 show time series of diffraction patterns recorded from the indicated positions. \textbf{(B-D)} Average diffraction patterns of the solid, hexatic, and liquid phases within the NBED map. The coloring in the NBED map is based on the approximate width of the diffraction spots: $\leq$ 0.20 rad $\rightarrow$ solid phase (blue); \textgreater~0.20 rad $\rightarrow$ hexatic phase (purple); isotropic rings $\rightarrow$ liquid phase (orange).}
    \label{fig:NBED}
\end{figure}

Although direct atomic-resolution imaging is ideal for probing static and sufficiently slow dynamics, it could only reveal averages of dynamics occurring on timescales shorter than the imaging process. Correspondingly, increasingly large parts of the area were washed out in the real-space images as the temperature was increased, and our information was drawn from the fluctuating sections where structures were still visible intermittently (see, for example, Movies S13 to S15). To complement our analysis, we used nano-beam electron diffraction (NBED) maps (Methods) to study a larger, though slightly irregular, AgI crystal shown in Fig.~\ref{fig:NBED}A. In the diffraction data, the time averaging takes place in reciprocal space, so that structural correlations can be probed even if the structure reorganizes rapidly, as in a liquid. In the example of Fig.~\ref{fig:NBED}, when the crystal was brought to  almost its final melting point, the left-hand side of the AgI flake was in a state that was dynamic on the time scale of single exposures (20 ms) and fluctuating between a solid and the phase that we identify as hexatic; Movie S4 recorded with a stationary probe at the position indicated in Fig.~\ref{fig:NBED}A shows the richness of these dynamics and that they were distributed both in time and space. The right-hand side of the crystal, in contrast, lost its hexagonal structure and continuous rings were visible in the diffraction patterns (Fig.~\ref{fig:NBED}D and Movie S5). These rings were static in the sense that they looked the same all over this area and did not change in subsequent exposures. We identified this phase as liquid, as detailed below.
The solid phase displayed sharp diffraction features (azimuthal width $\leq$ 0.20 rad) of a hexagonal lattice. We attribute the asymmetry in the intensities to a varying level of sample tilt as in Figure~\ref{fig:NBED}B (see also Figure \ref{fig:diffraction_tilted} and our previous work~\cite{mustonen_toward_2022}). In the hexatic phase, in contrast, we observed azimuthally broadened diffraction peaks (Figure~\ref{fig:NBED}C). This azimuthal broadening is consistent with a lattice that is distorted by the dynamic appearance and annihilation of dislocations or grain boundaries \cite{toledano_melting_2021}. Although the orientational order was clearly reduced compared to the crystalline phase, it was nevertheless present and maintained across the entire portion of the crystal that had not transitioned to the liquid state.

The spatial correlation length in liquids is much shorter than in crystalline materials and typically extends no further than a few nearest neighbors. In our diffraction patterns, the complete loss of orientational order was evident from the ring-shaped intensities, which also indicated that the structure was dynamic on a time scale much faster than the diffraction pattern acquisition time. Nevertheless, some insights on the near-ordering could be gained from the radial profile (Fig.~\ref{fig:NBED}D): The primary distances were the same as for crystalline AgI, suggesting that the building blocks and bonding in the liquid phase were the same as in the solid. Compared to crystalline AgI, radial broadening of the peaks was observed, especially in the second-order diffraction ring of the liquid phase, but also in the measured width of the first-order ring (Figure~\ref{fig:line_profiles}). This broadening corresponded approximately to a correlation length between 0.6 and 2 nm, as discussed in the Supplementary Material.

These observations were not just in excellent qualitative agreement with the hexatic phase, but as we will show below, also in quantitative agreement. The hexatic phase can be rigorously defined by how the translational and orientational orders are correlated as a function of distance within the surrounding material. The classical definition expects an exponential decay of the translational correlation within the hexatic phase, distinguishing it from the power-law decay expected in solids at a finite temperature. The hexatic state retains quasi-long-range orientational order, distinguishing it from both the crystalline phase and the liquid phase. A fully crystalline phase has a long range orientational order whereas a liquid phase has only a short range order. In the KTHNY theory these conditions can be expressed in terms of specific critical exponents. The hexatic phase must simultaneously display a translational correlation power-law decay exponent $\eta_k\ge 1/3$ and an orientational correlation decay exponent $\eta_6\le 1/4$~\cite{bedanov_modified_1985, zahn_two-stage_1999, bernard_two-step_2011}. These decay characteristics are seen in Fig.~\ref{fig:Correlations}, as we shall discuss in the following.

\section*{Translational correlation}

\begin{SCfigure}[1.0][!t]
    \centering
   \includegraphics[width=0.67\textwidth]{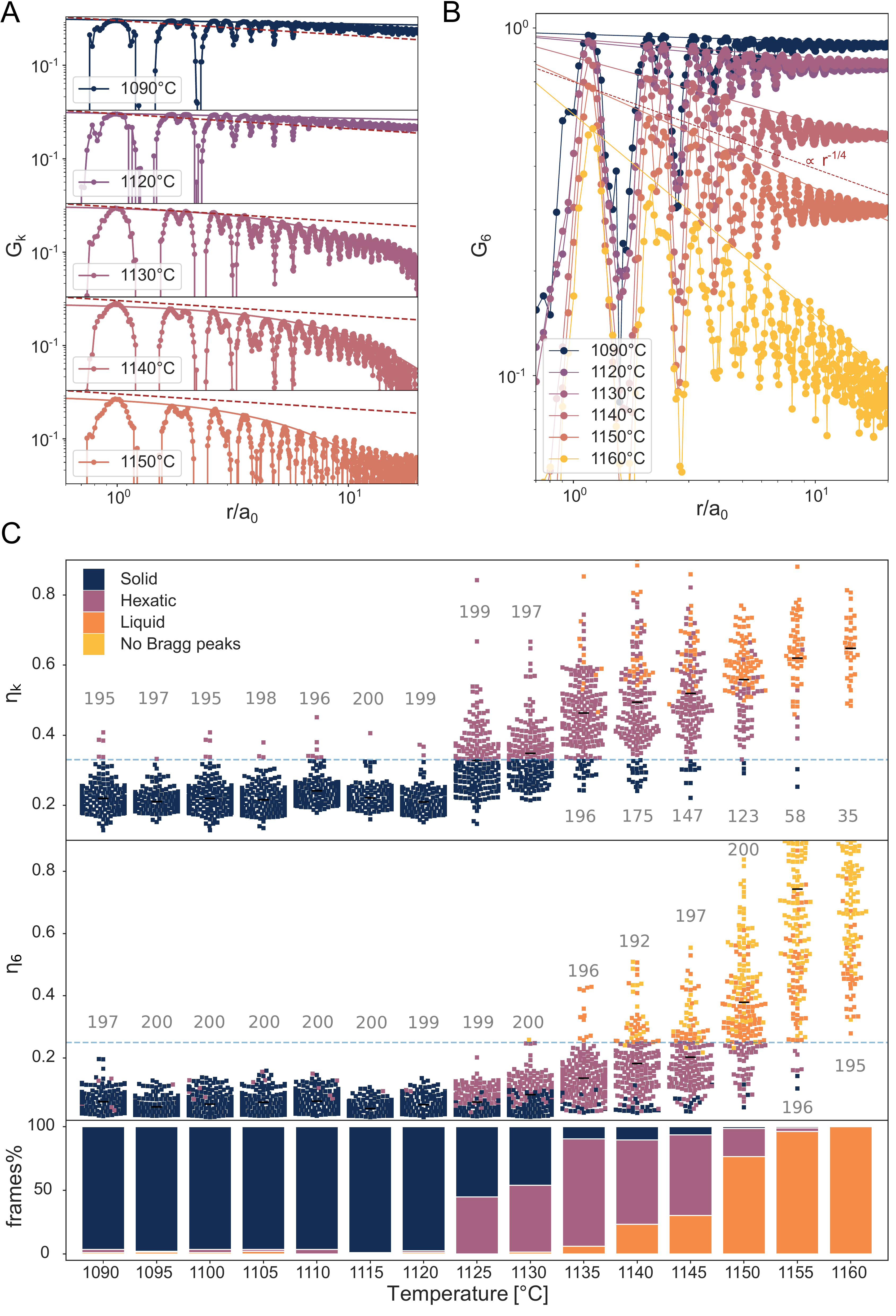}
    \caption{\textbf{Spatial correlation functions.} \textbf{(A)} The average translational correlation function $G_k$ at various temperatures with solid lines indicating either power-law fits (1090~\degree C and 1120~\degree C) or exponential fits (1130~\degree C, 1140~\degree C, and 1150~\degree C). The critical exponent $\eta_k \rightarrow 1/3$ is indicated with brown dashed lines. \textbf{(B)} The average orientational correlation function $G_6$ with solid power-law fits. The critical exponent $\eta_6 \rightarrow 1/4$ is plotted with brown dashed lines. As negative values cannot be plotted in a log-log coordinate system, some data points in (A) and (B) are not visible. \textbf{(C)} Fit parameters for $G_k(r)$ and $G_6(r)$ as a function of temperature for every image. Data are classified as: solid phase ($\eta_k \leq 0.33$), hexatic phase ($\eta_k > 0.33$ and $\eta_6 < 0.25$), and liquid phase ($\eta_6 > 0.25$). Relative phase abundances are shown below. All images lacking reciprocal lattice vectors are considered liquid. The number above or below the individual columns indicates the number of analyzed frames.}
    \label{fig:Correlations}
\end{SCfigure}

We computed the translational correlation functions using the center of mass positions $r$ of the AgI polygons detected automatically in real space STEM ADF images by the CNN (see Supplementary Figure~\ref{fig:CNN} and Movies S6-S15). Locally, the deviation from the perfect hexagonal crystal can be obtained by computing the translational order parameter
\begin{equation}
    \Psi_l(r) = e^{-iq_l r},
\end{equation}
where $q_l$ is a reciprocal lattice vector of a perfect crystal. The reciprocal lattice vectors were found by first computing a 2D structure factor $S(q)$~\cite{ramasubramani_freud_2020} and then fitting the first-order peaks (see Supplementary Material Figure~\ref{fig: SF} for examples). The translational correlation function was then computed using $\Psi_l$ as
\begin{equation}
    G_k(r) = \frac{1}{6}\sum_{l = 1}^6\frac{1}{N_r}\sum_{i,j}^{N_r} \Psi_l(r_i)\Psi^*_l(r_j).
    \label{equ: Translational}
\end{equation}
The correlation function was averaged over $N_r$ pairs of polygons, and over the first six reciprocal vectors in $S(q)$.
Fig.~\ref{fig:Correlations}A shows $G_k$ averaged over all images at each temperature together with the best fitting power-law $\propto$ $r^{-\eta_k}$ and exponential $\propto$ $e^{-r}$ decay functions. The statistical distribution of the power law fitting parameter $\eta_k$ of all recorded images can be found in Fig.~\ref{fig:Correlations}C.

Power-law decay of $G_k$ $\propto$ $r^{-\eta_k}$ with low $\eta_k$ was observed in nearly all images at 1090~\degree C, as expected for a solid system. As the temperature was raised the power law fitting exponent remains nearly constant and well below the critical value of $\eta_k = 1/3$ required by the definition of the hexatic phase until 1120~\degree C. However, at 1125~\degree C the decay of the averaged translational correlation function suddenly became quasi-exponential and at 1130~\degree C $\eta_k$ exceeded 1/3 in more than one half of the recorded images as the material fluctuated between crystallinity and the hexatic state, as shown in (Movie S9).

With increasing temperature, the average $\eta_k$ rose steadily until the highest recorded temperature of 1160~\degree C, where all images had an associated translational decay parameter \textgreater1/3, indicating a complete loss of translational order with an exponentially falling average $G_k$. As $G_k$ decays exponentially in both the hexatic and liquid phase, power-law fits were not strictly applicable, leading to a wide spread in $\eta_k$. However, the key point is that the decay parameter remains greater than the critical 1/3 value in most recorded images, implying the presence of either the hexatic phase or a semi-ordered liquid. At 1160~\degree C, two-thirds of the images had no discernible first-order diffraction spots, indicating that the 2D crystal had almost completely transformed to the liquid state.
\clearpage
\section*{Orientational correlation}

To distinguish the hexatic phase from the solid and liquid phases, we must also assess the orientational order, which is determined by analyzing the orientational correlation
\begin{equation}
   G_6(r) = \frac{1}{N_r}\sum_{i,j}^{N_r} \Psi_6(r_i)  \Psi^*_6(r_j)
\end{equation}
in which the local orientational order parameter is defined as
\begin{equation}
    \Psi_6(r_i) = \frac{1}{N_{nn}}\sum_{j=1}^{N_{nn}} e^{i6\theta_{ij}},
\end{equation}
where $\theta_{ij}$ is the angle between a polygon at position $r_i$ and its $N_{nn}$ nearest neighbors relative to a fixed axis. The nearest neighbors were determined by Voronoi tessellation~\cite{ramasubramani_freud_2020}. Figure~\ref{fig:Correlations}B shows $G_6$ averaged over all images at each temperature and the corresponding fits on a log-log scale. The statistics of the power law fitting parameter $\eta_6$ among all images can be found in Fig.~\ref{fig:Correlations}C. 

During the heating experiments, $G_6$ remained nearly constant up to 1120~\degree C as expected for a solid. However, as the temperature was increased to 1125~\degree C, the orientational correlation function entered into a power law decay $\propto$ $e^{-\eta_6}$, corresponding to a linear decrease on the log-log scale. However, the decay parameters, $\eta_6$, remained low and did not exceed the critical value $\eta_6 = 1/4$ for the hexatic-liquid transition until 1145~\degree C. Nevertheless, even at 1145~\degree C, over 75\% of the images displayed orientational decay consistent with the hexatic phase, i.e., with $\eta_6 < 1/4$. Both the orientational order and translational order correlations agreed with the definition of the hexatic phase between approximately 1125 and 1145~\degree C.

At 1150~\degree C a sudden increase in $\eta_6$ occured, accompanied by a large deviation in the fitting parameter values, consistent with an increasingly liquid character of the crystal. At 1155~\degree C the orientational correlation vanished completely, and at 1160~\degree C the AgI crystals could be considered molten with an average $\eta_6$ of 1.12. These observations are in excellent agreement with the FTs of the datasets shown in supplementary Figure~\ref{fig:FT_Figure3}, which displayed a gradual azimuthal broadening of the AgI peaks at temperatures above 1125~\degree C. Note, however, that the power-law-like decay of $G_6$ of most frames at the highest temperatures stands in contradiction to the KTHNY theory. As this decay is not represented in every frame it is unlikely that it arose from an artifact of the CNN detection at high temperatures. Instead, it may be a manifestation of a melting scenario that is of a non-KTHNY type.

\section*{On the melting mechanism}
The translational and orientational correlation functions revealed the presence of a hexatic phase in a narrow temperature window below the melting point, but these parameters alone do not determine the melting mechanism. According to the KTHNY theory, both transitions — solid to hexatic and hexatic to liquid — are continuous, whereas the mixed melting scenario features a continuous solid to hexatic transition followed by a first-order hexatic to liquid transition \cite{toledano_melting_2021, chui_grain-boundary_1982, chui_grain-boundary_1983}. 

\begin{figure}[!b]
    \centering  \includegraphics[width=0.9\textwidth]{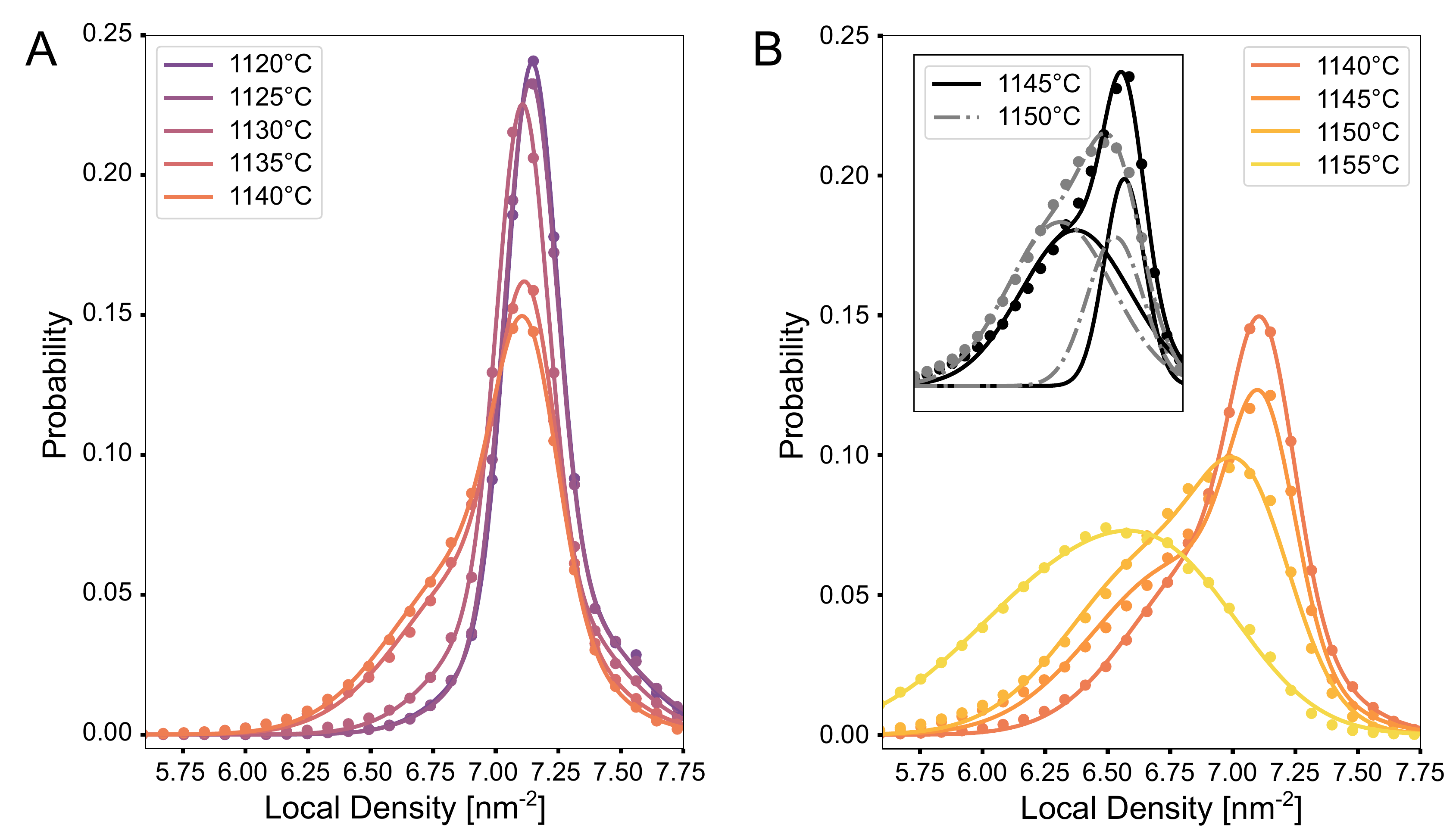}
    \caption{\textbf{Local density distribution functions at different temperatures.}~\textbf{(A)} Change of local density in the solid-hexatic transition temperature range with Gaussian and double Gaussian fits. ~\textbf{(B)} Change of local density in the hexatic-liquid transition temperature range with Gaussian and double Gaussian fits. The inset contains the deconvolution of the double Gaussian fits of the 1145~\degree C and 1150~\degree C datapoints, showing the bimodal nature of these distributions.}
    \label{fig:density_distribution}
\end{figure}

In continuous melting, local order evolves gradually with temperature, driven by the unbinding of defect pairs. In contrast, a first-order transition results in an abrupt change in local order and the coexistence of two distinct phases, typically separated by grain boundaries. To distinguish between these transition types, the local density distribution of the particles is a useful indicator, exhibiting a bimodal character when two phases coexist. Figure \ref{fig:density_distribution} illustrates how, in our experiment, the local density evolves across the solid–hexatic and hexatic–liquid transitions.
Apart from minute broadening and a subtle shift toward lower density, the density distribution remained nearly unchanged and unimodal up to 1130~\degree C, consistent with a continuous solid–hexatic transition. However, starting at 1135~\degree C, the distribution became increasingly bimodal and by 1150~\degree C, the lower-density mode became dominant. These changes provide evidence for phase coexistence and indicate a first-order hexatic-to-liquid transition near the melting point, consistent with a mixed melting scenario \cite{toledano_melting_2021}. At 1155~\degree C, the local density distribution became very wide and quasi-unimodal, indicating the complete loss of correlation in the liquid phase.

In the KTHNY scenario, melting is driven by the formation of high-energy (charged) dislocation defects. The six-membered rings of thy AgI lattice can deform into 5-7 pairs (dislocations) at the solid-hexatic-transition, which then unbind into isolated 5- and 7-membered rings (disclinations) at the hexatic-liquid transition. This leads to structural disintegration even at low defect densities. In contrast, the mixed melting scenario involves lower energy defects that are more abundant and rapidly form large clusters, and thus create grain boundaries between different coexisting phases~\cite{toledano_melting_2021}.

We analyzed the abundance of different defect types and clusters in our system through Voronoi segmentation. As the amount of strictly isolated dislocations and disclinations remained low at all temperatures, we classified the symmetry-breaking effect of defect clusters by identifying their topological charge. Defect clusters were classified into dislocation, disclination, and neutral-charged categories, as well as into an "unknown" category for clusters that could not be classified (see discussion in Supplementary Material and Fig.~\ref{fig: Defects}). The abundance of defects inside both dislocation and disclination charged clusters remained constant up to 1125~\degree C, after which both types of defects increased rapidly and continuously. At higher temperatures, a substantial number of defects began to form line-like clusters, which rapidly grew outside the observed field of view, complicating the determination of the type of defect and leading to a fast increase in the "unknown" category. The high defect density at the highest temperatures, where the ratio of topological defects (non-hexagonal Voronoi cells) to hexagonal cells reached 0.35, and the absence of isolated dislocation unbinding coupled with the observation of a first-order hexatic-liquid transition supported a mixed-type melting scenario consistent with the grain boundary theory of melting \cite{saito_monte_1982}. 

\section*{Conclusions}

We have recorded the phase transitions from solid to liquid in a covalently bonded 2D crystal, and demonstrated that melting proceeds via a hexatic intermediate phase. Graphene encapsulation enabled us to observe this transition in 2D AgI through time- and temperature-resolved in-situ atomic-resolution STEM imaging and NBED. The hexatic phase, which forms only in 2D systems, exhibited long-range orientational but only short-range translational order, and emerged within a 25~\degree C window below the melting point. The absolute transition temperature depended on crystal size. Although a definitive distinction between the KTHNY and mixed-melting scenarios, both involving a hexatic phase, remains challenging, our data suggest that 2D AgI followed a mixed-melting pathway. These findings provide direct insight into 2D melting in a real covalent crystal and open new directions for understanding phase transitions in atomically thin materials.

\bibliography{ref}

\bibliographystyle{Science}

\section*{Acknowledgments}
We thank Xian-Bin Li and Dan Wang for kindly providing the AgI vacancy model structures. JCM declares an additional affiliation at the NMI Natural and Medical Sciences Institute at the University of Tübingen.

\section*{Funding}
KM and TAB acknowledge funding from FWF through a grant number P35912. VS has been supported by the V4-Japan Joint Research Program V4-Japan/JRP/2021/96/BGapEng, the Grant No. VEGA 1/0104/25, provided by Ministry of Education, Research, Development and Youth of the Slovak Republic, the EU NextGenerationEU through the Recovery and Resilience Plan for Slovakia under the project No. 09I05-03-V02-00071 and the QM4ST, CZ.02.01.01/00/22-008/0004572 of Johannes Amos Commenius Programme for Excellent Research. T.J.P. acknowledges support from the European Research Council (ERC) under the EU Horizon 2020 program, Grant Agreement No. 802123-HDEM. LF and DL acknowledge funding from FWF through grant number P35318 and DOC142. M.K. acknowledges funding from the National Center for Research and Development (NCBR) under the V4-Japan Joint Research Program BGapEng V4-JAPAN/2/46/BGapEng/2022.

\section*{Authors Contributions}
KM and TAB conceived the study. TAB, KM, DL, CM, and VS conducted the experimental work, while DL, KM, TAB, and JCM performed the data analysis. JM developed and trained the neutral network. PK and VS synthesized the sample material, and MK and AM provided computational support. KM, TJP, DL, JCM, TAB, VS, JK, and LF co-wrote the manuscript.

\section*{Competing interests}
The authors declare no competing interests.

\section*{Data and material availability}
The experimental data supporting these findings are openly available in the University of Vienna repository PHAIDRA at https://doi.org/10.25365/phaidra.718 \nocite{bui_data_2025}.

\section*{Supplementary Materials}
\noindent Materials and Methods

\noindent Supplementary Text

\noindent Figures S1 to S30

\noindent Supplementary Movies S1 to S15

\noindent References 39 - 61


\renewcommand{\appendixname}{Supplementary Material}
\setcounter{page}{2}
\renewcommand{\thefigure}{S\arabic{figure}} \setcounter{figure}{0}
\renewcommand{\thetable}{S\arabic{table}} \setcounter{table}{0}
\renewcommand{\theequation}{S\arabic{table}} \setcounter{equation}{0}

\section*{Materials and Methods}

\subsection*{Sample Preparation}

\begin{figure}[ht]
    \centering
    \includegraphics[width=\textwidth]{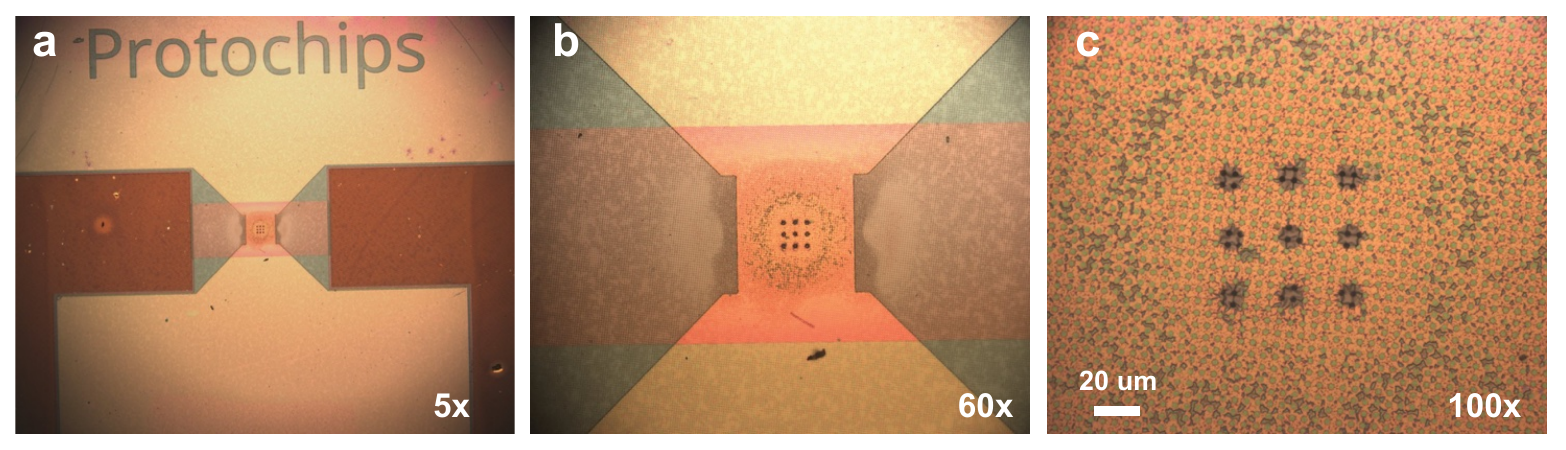}
    \caption{\textbf{Optical compound microscope images of a Protochips Fusion heating holder.} The graphene encapsulated 2D AgI is suspended on an amorphous carbon film, which is supported on a silicon nitride/silicon carbide membrane. The carbon membrane is only visible in panel \textbf{c}, whereas the silicon nitride/silicon carbide membrane with a total of nine holes, is visible in the lower magnifications images in \textbf{a-b} as well.}
    \label{fig: protochips}
\end{figure}

The heterostructures of graphene-encapsulated AgI were synthesized using the chemical process detailed in our previous work, where silver nitride served as the metal precursor and iodic acid as the iodine source \cite{mustonen_toward_2022}. The structures were grown directly on Protochips Fusion heating holders, which were subsequently used for the \textit{in situ} heating experiments. The freestanding portions of the sample were supported by a holey carbon film  with a thickness of ca. 10 nm, mounted on a 200 nm thick silicon nitride/silicon carbide membrane with 10 $\mu$ perforations, as shown in Figure \ref{fig: protochips}.

\subsection*{Scanning Transmission Electron Microscopy (STEM)}

All microscopy data in this study were acquired using an aberration-corrected Nion UltraSTEM 100 scanning transmission electron microscope. The experiments were performed at an acceleration voltage of 60 kV and a convergence semiangle of ca. 35~mrad, and the images captured using a high-angle annular dark-field (HAADF, cited in the manuscript as ADF) detector featuring a collection semi-angle of 80–300 mrad. The electron energy loss spectra in Figure \ref{fig: EELS} were acquired using a custom-built device comprising a Gatan PEELS 666 spectrometer and an Andor iXon 897 electron-multiplying charge-coupled device (EMCCD) camera with 0.5~eV/px energy dispersion. The Protochips Fusion sample carriers (see above) were resistively heated by a Keithley sourcemeter model 2614B controlled through Protochips Fusion proprietary software. The accuracy of the temperature setpoint provided by the manufacturer is $\pm$ 5\%. In these experiments a temperature ramping rate of 6~\degree C/min was used, and a 15-min thermal stabilization period was allowed before data collection at each temperature setpoint.

\subsection*{STEM Acquisition Parameters}

The overviews showing the AgI crystal in Figure \ref{fig:2D_melting} were acquired with a probe position dwell time of 8.0 $\mu$s using a 2048 $\times$ 2048 probe position array. The electron beam fly-back time was 120~$\mu$s per scan line in all experiments, in this case resulting in a total image acquisition time of 33~s. The Fourier transforms shown in Figure~\ref{fig:2D_melting} were averaged from Fourier transforms of 100 individual ADF images. Each of these images consisted of a 1024 $\times$ 1024 probe position array with a dwell time of 0.5~$\mu$s. This resulted in a total image acquisition time of 0.65~s. The images used for the spatial correlation function analysis (Figure \ref{fig:Correlations}) were acquired with a dwell time of 1.0~$\mu$s and a 1024 $\times$ 1024 probe position array, resulting in a acquisition time of 1.17~s per image. A total of 200 images were captured for each temperature condition. The nano-beam electron diffraction (NBED) patterns (Figure~\ref{fig:NBED} and Supplementary Videos SV4 and SV5) were acquired using a nearly parallel beam illumination mode with an approximate probe size of 5~nm. Each diffraction pattern was recorded with an acquisition time of 20~ms. For the NBED map, a 40 $\times$ 40 probe position array was used, with a step size of 1.75~nm. A background subtraction for the NBED images in Figures~\ref{fig:NBED} and \ref{fig:line_profiles} was performed by averaging 20 NBED images taken outside the crystal and subtracting this contribution from each diffraction pattern; this is also the reason why the graphene reflections near the image perimeter appear as dark spots.

\subsection*{Time-Averaged Fourier Transforms}
\begin{figure}[b!]
    \centering
    \includegraphics[width=0.9\textwidth]{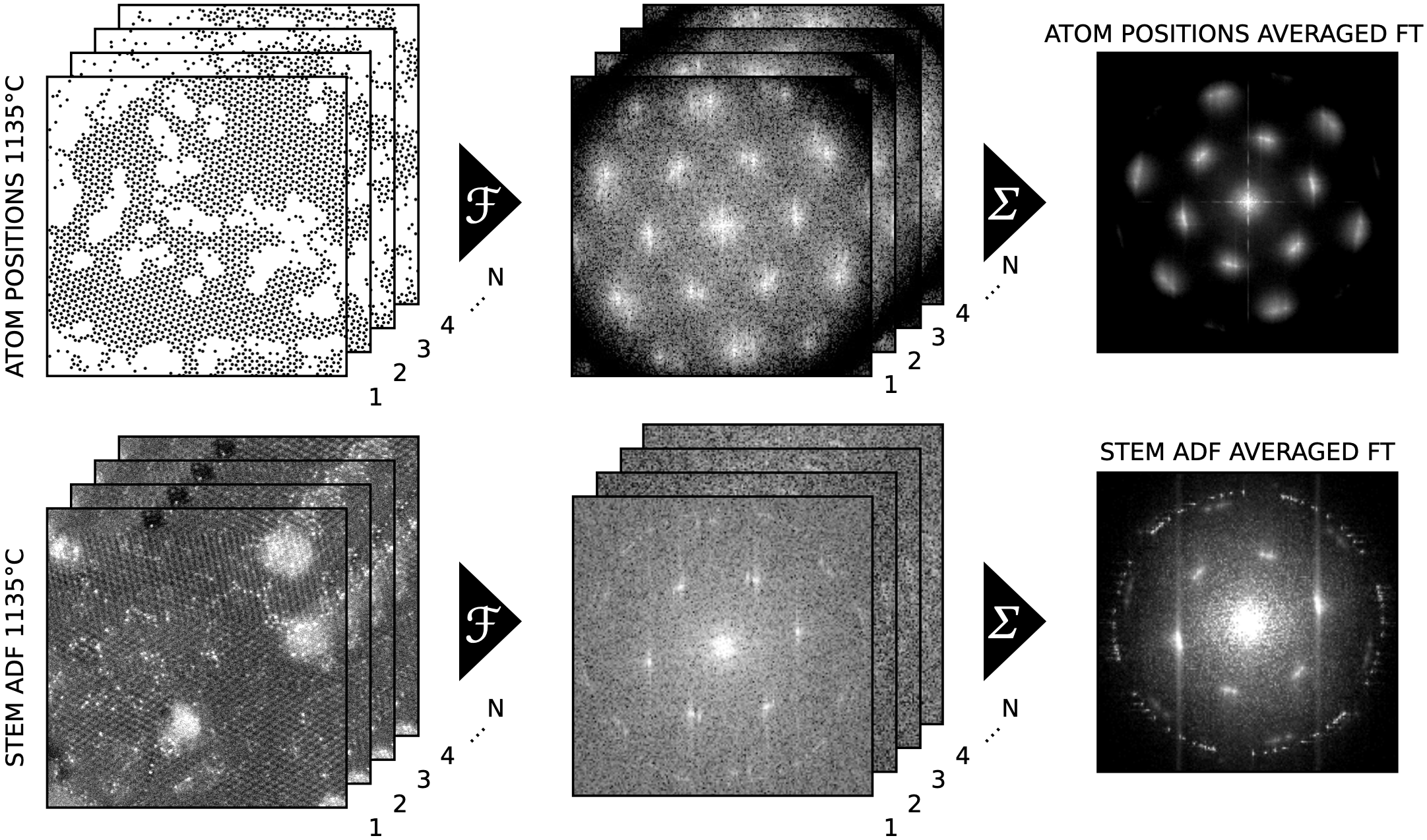}
    \caption{\textbf{Time-averaged Fourier transforms} Workflow for obtaining time-averaged Fourier transforms of STEM-ADF images and CNN-extracted atom positions.}
    \label{fig: FT_method}
\end{figure}

The time-averaged Fourier transforms shown in Figures \ref{fig:2D_melting} and \ref{fig:FT_Figure3} were generated using the ImageJ software suite. This process involved computing a 2D FT for each STEM ADF image (16 nm $\times$ 16 nm each) of an image stack containing either 100 (Figure \ref{fig:2D_melting}) or 200 (Figure \ref{fig:FT_Figure3}) individual frames, and subsequently averaging the resulting transforms into time-averaged FTs. This workflow is schematically presented in Figure \ref{fig: FT_method}. In addition, to evaluate the reliability of the CNN output, we also computed FTs based on the detected atom positions (see Figure \ref{fig:CNN}) extracted from the STEM-ADF frames at temperatures around the phase transitions, and compared these to the FTs of the original STEM ADF datasets. These results are compared in Figure \ref{fig:CNN_FT}.
\subsection*{STEM Simulations}
The simulated ADF images in Figures \ref{fig:retrieved}, \ref{fig:CNN_distortions}, and \ref{fig:tilt} were computed using the \textit{ab}TEM code~\cite{madsen_abtem_2021}. The atomic models used for the simulations were produced either via density functional theory (DFT) calculations or a convolutional neural network (CNN, see below). All simulations employed realistic instrument parameters: 60 kV acceleration voltage, 80–300 mrad ADF collection semiangle, and 35 mrad probe convergence semiangle. For DFT structures, potential and probe sampling were set to 0.03~\AA, using Nyquist sampling for the grid scan, while CNN models used 0.1~\AA~potential and probe sampling with Nyquist sampling of 0.358~\AA. Poisson noise was applied to the simulated images, replicating the experimental electron dose per frame of 4.2 × 10$^4$ electrons/\AA$^2$ for Figure \ref{fig:tilt}, 200 × 10$^3$ electrons/\AA$^2$ for Figure \ref{fig:iodine_STEM}, and 1.3 × 10$^4$ electrons/\AA$^2$ for all other figures and data.

\subsection*{Density Functional Theory Modeling}

First principles calculations were performed using the plane wave Q{\sc{uantum}} ESPRESSO (QE) package~\cite{QE-2009,Giannozzi_2017}. We used the Perdew–Burke–Ernzerhof exchange-correlation functional ~\cite{perdew_restoring_2008} as implemented in the full-relativistic SG15 optimized norm-conserving Vanderbilt (ONCV) pseudopotentials~\cite{Hamman2017,Scherpelz2016}. The plane wave basis was converged using 
the kinetic energy cut-offs 60\,Ry and 240\,Ry for the wave function and charge density, respectively. The two-dimensional heterostructure was modeled by introducing a vacuum 17.8~\AA~in the stacking (z-) direction minimizing the interaction between the periodic images of the system. Van der Waals interactions were taken into account within the semi-empirical Grimme's DFT-D2 corrections \cite{grimme2006,barone2009}. Self-consistency was achieved using $10\times 10\times 1$ Monkhorst-Pack \cite{MPack} $k$-point mesh for the supercell calculations. The optimized lattice constant of AgI, $a_{\text{AgI}}=4.59$ \,\AA, was found by variable cell relaxation, assuming force and total energy thresholds $10^{-4}$\,Ry/bohr and $10^{-5}$\,Ry respectively, and  allowing the cell to move in the two-dimensional plane.

\section*{Obtaining the Atomic Structures from the ADF Images}

\subsubsection*{The Architecture of the Convolutional Neural Network (CNN)}

The CNN used in this work was designed for analyzing microscopy images features a UNET architecture with rotational equivariance, based on the approach developed by Weiler \textit{et al.} \cite{weiler_general_2019}. Similar architectures have been widely adopted for analysis of atomic structures and strain directly from noisy STEM images \cite{lee_deep_2020, lee_stem_2022, ziatdinov_deep_2017, kalinin_machine_2022, trentino_atomic-level_2021}. Our neural network is specifically trained to detect atoms in simple hexagonal lattices in images with a pixel size of 0.1~\AA/px, and images with other pixel sizes are rescaled accordingly. The scaling factor is determined directly from the known AgI lattice vectors using a robust automated algorithm in Fourier space. The image mean and standard deviation are normalized, but in images with significant contamination, the statistics can be heavily skewed due to the contrast difference between the lattice and contamination regions. To address this, we apply the neural network in two stages: in the first pass, we generate a segmentation map to create a mask that excludes heavily contaminated areas from the statistics. Note however that high-contrast contamination areas in our data still in many cases contain sufficient information to interpret the most likely atom positions, as for instance in Figure \ref{fig: CNN_contamination}. The second pass then uses these contamination-masked images with improved normalization for the final prediction. While this process could be iterated until convergence, we find that two passes are typically sufficient. The full description of the CNN architecture used for this work can be found in Ref. \citenum{trentino_atomic-level_2021}.

\begin{figure}[!t]
    \centering
    \includegraphics[width=12cm]{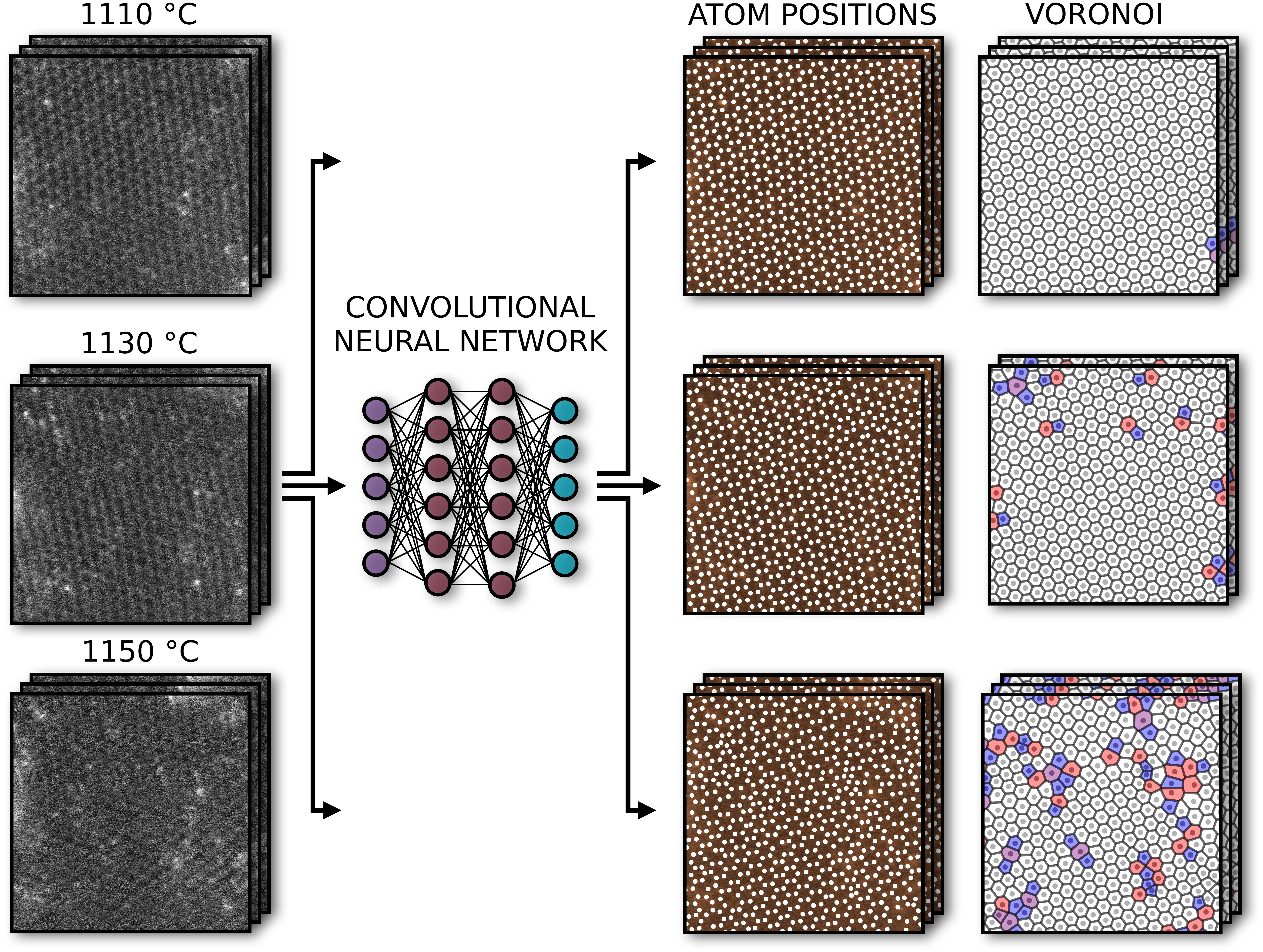}
    \caption{\textbf{Segmentation workflow.} STEM ADF input stacks are analyzed automatically using a convolutional neural network (CNN) to detect the most probable atom positions in each frame. To study the defect configurations and polygon orientations, the polygon center of mass is first determined using the stable Delaunay algorithm, and the configurations are segmented further by applying Voronoi tessellations. The example STEM images have a field of view of 7.4 nm × 7.4 nm.}
    \label{fig:CNN}
\end{figure}

\begin{figure}[!ht]
    \centering
    \includegraphics[width=12cm]{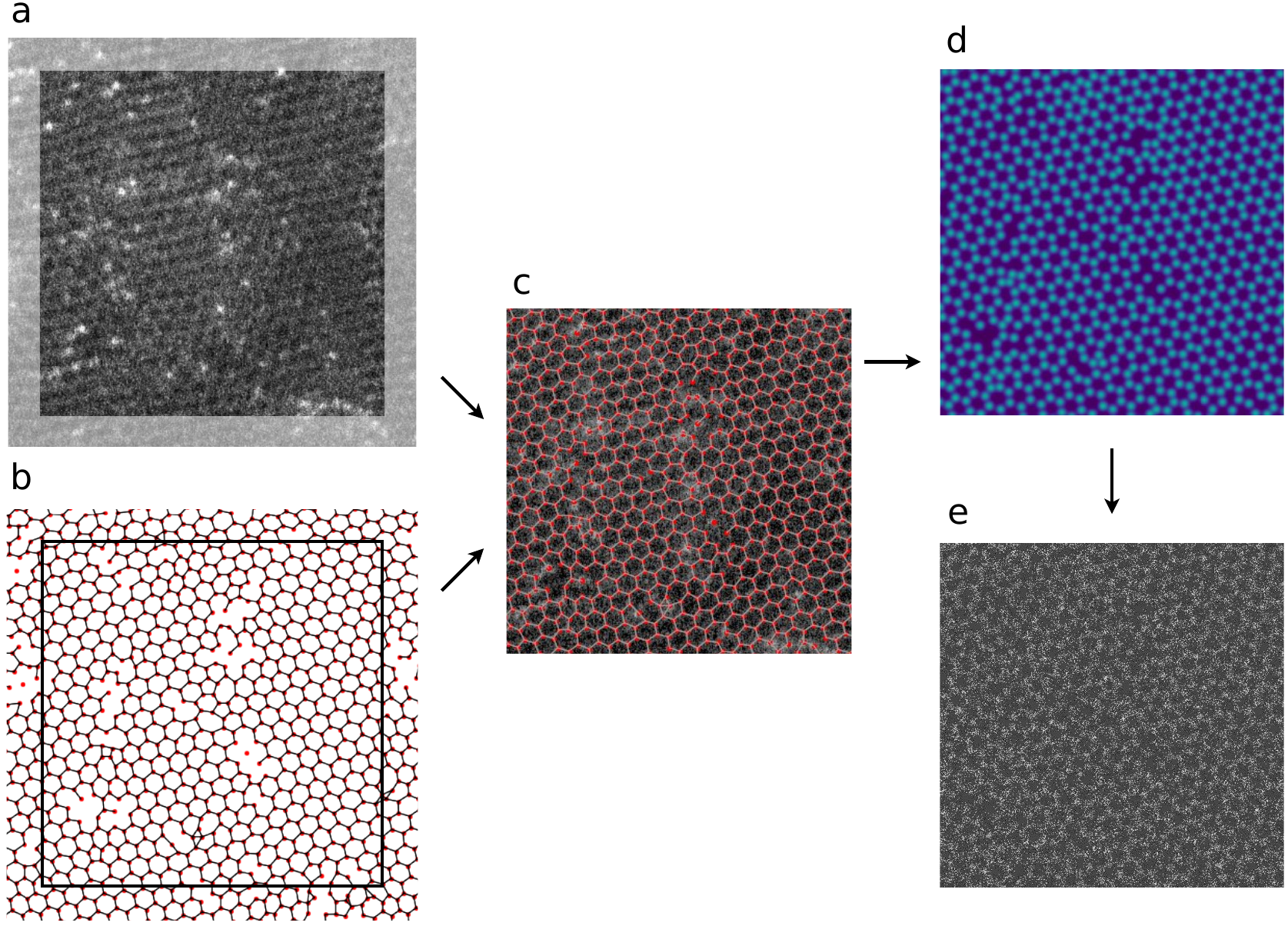}
    \caption{\textbf{Comparison of the extracted atomic structure and the original STEM image.} \textbf{(a)} Original STEM ADF image. \textbf{(b)} Delaunay wire mesh based on the CNN output with atom positions marked with red dots. \textbf{(c)} The original STEM image overlaid with the wire mesh. Simulated STEM ADF image of the extracted atomic structure is shown in \textbf{(d)} without noise and with a $\sigma$ = 0.6 Gaussian filter, and in \textbf{(e)} with realistic acquisition-related Poisson noise.}
    \label{fig:retrieved}
\end{figure}

In our workflow (Figure \ref{fig:CNN}), the CNN output is first converted into a point map representing all discernible atoms within the image area. For validation, this atom position map can also generate a simulated image of the most likely initial structure, as shown in Figure \ref{fig:retrieved}. The stable Delaunay algorithm is then applied to determine the center of mass of each polygon defined by the atoms, which is subsequently used for Voronoi tessellation \cite{ramasubramani_freud_2020}, as shown on the right in Figure \ref{fig:CNN}. In the fitting process, we allow the Delaunay algorithm to resize and translate the polygons without skewing them, ensuring high reliability for determining the center of mass of the polygon independent of minor sample drift and other minor image distorting factors. The spatial correlation functions are computed from the tessellated region that excludes the edges.

\subsubsection*{Benchmarking the CNN}

Several factors can impact the detection accuracy of the CNN. These factors include at least electron beam-induced and dose-related structural changes in the studied material, image noise related to the thermal instability of the sample as well as to the detector noise, and sample contamination, consisting primarily of amorphous material on the outer surfaces of the encapsulating graphene layers. These factors and their influence on our analysis are now discussed.

\begin{figure}[t!]
    \centering
    \includegraphics[width=0.90\textwidth]{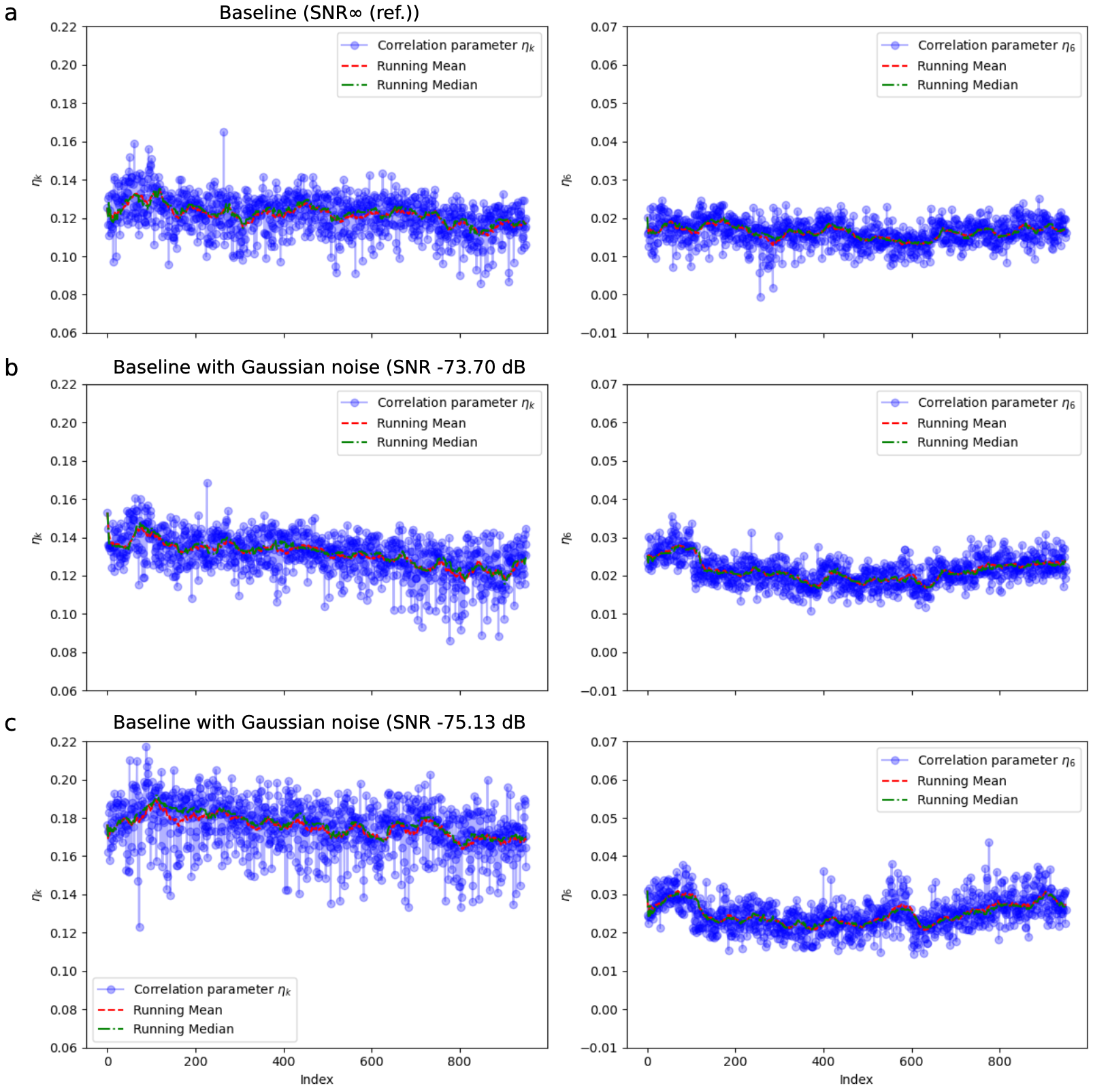}
    \caption{\textbf{Baseline of the spatial correlation function decay exponents.} \textbf{a} Translational ($\eta_k$) and orientational ($\eta_6$) decay exponents calculated for 952 rapidly acquires STEM ADF images at room temperature. \textbf{b-c}  The correlation function decay exponents computed for the same data with artificial Gaussian noise (see Figure \ref{fig: CNN_noise}).}
    \label{fig: CNN_baseline}
\end{figure}

\begin{figure}[t!]
    \centering
    \includegraphics[width=0.60\textwidth]{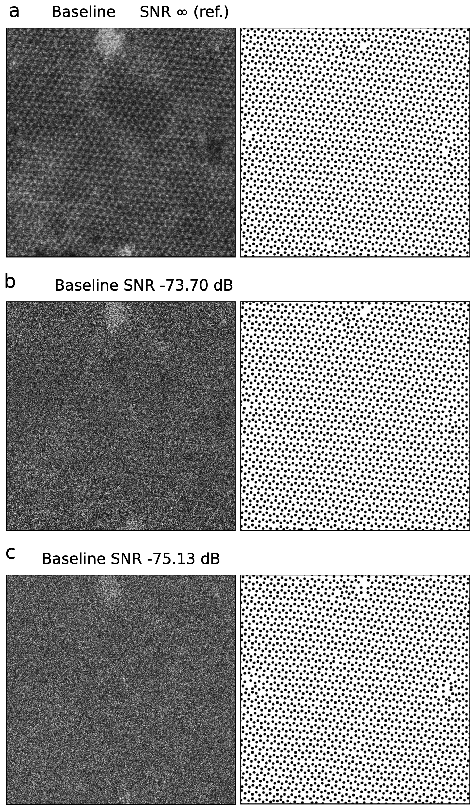}
    \caption{\textbf{Snapshots of the baseline data.} \textbf{a} The first frame of the baseline dataset with CNN-detected atom positions. \textbf{b-c} The same image with added artificial Gaussian noise. \textbf{d} Translational ($\eta_k$) and orientational ($\eta_6$) decay exponents calculated for 952 images, both with and without added Gaussian noise.}
    \label{fig: CNN_noise}
\end{figure}

\paragraph{Electron Beam Damage} Electron beam damage can occur through knock-on displacement, where atoms are ejected from the lattice due to direct momentum transfer, or through radiolysis, which induces bond breaking and vacancy formation via inelastic energy transfer \cite{susi_quantifying_2019}. Vacancies, if created, would eventually perturb the crystal structure and cause its spatial correlations to decay faster as the electron dose accumulates. On the other hand, graphene encapsulation is known to mitigate such damage by replenishing electrons and confining displaced atoms, allowing them to recombine with the created vacancies \cite{zan_control_2013}. The shielding effect of Graphene is also evident in our ``baseline" dataset, which was acquired to simulate the temporal evolution of spatial correlations unrelated to heating during extended heating experiments. These results, which include the spatial correlation decay exponents for 952 continuously acquired STEM-ADF frames, are summarized in Figure \ref{fig: CNN_baseline}a, with a snapshot of the dataset shown in Figure \ref{fig: CNN_noise}a. The minimal observable change in the decay exponents confirms that electron irradiation alone does not induce irreversible changes in graphene-encapsulated AgI. The accumulated electron dose during the baseline data collection was approximately $1.45 \times 10^7$ electrons/\AA$^2$, with a total data collection time of around 1114 s.

\paragraph{CNN Handling of Thermal Vibrations} Random noise, arising from thermal instabilities of the sample and rapid AgI structural reordering, introduces errors in CNN-based atom position estimation and thus contributes to uncertainty in computing spatial correlation functions. We evaluated the noise tolerance of the CNN by adding artificial Gaussian noise on the baseline STEM-ADF images (see Figure \ref{fig: CNN_noise}a), and then reevaluated the correlation decay exponents frame-by-frame. These results are summarized in Figure \ref{fig: CNN_baseline}b–c, while example snapshots of noise-added copies of the same frame are shown in Figure \ref{fig: CNN_noise}b–c. These data show that while the translational decay exponents ($\eta_k$) are slightly higher for noisy input data, the overall contribution is not significant even in case of extremely noisy input with a SNR of -75.13 dB. The SNR for the noisy STEM ADF images was here calculated as
\begin{equation}
 SNR = 10 \cdot \log_{10} \left( \frac{{\mathrm{Signal~Power}}}{\mathrm{Noise~Power}} \right),
 \tag{Eq. S1}\label{equ: SNR}
\end{equation}
where $\mathrm{Signal~Power}$ is the variance (or mean squared value) of a reference image (e.g. Figure \ref{fig: CNN_baseline}a), and $\mathrm{Noise~Power}$ is the variance of the difference between the noisy image and the reference.

\paragraph{CNN Handling of Contamination} In some of our ADF data bright areas are visible that are caused by electron scattering from amorphous material on the graphene surfaces. These features, originating from sample fabrication and accumulating during electron beam exposure, often obscure the underlying AgI lattice from a human observer, as image contrast cannot be adjusted to simultaneously reveal both low- and high-brightness areas. 
So as long as sufficient structural information is present the CNN infers the likeliest atom positions even underneath the contaminated areas. 
 
\begin{figure}[!ht]
    \centering
    \includegraphics[width=12cm]{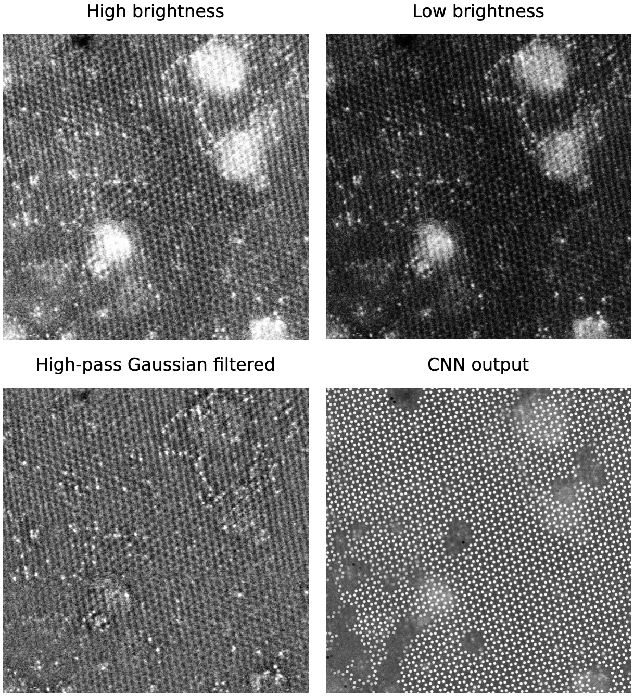}
    \caption{\textbf{CNN handling of amorphous sample contamination.} The high brightness contamination areas in STEM-ADF images (top left) result in low contrast of the AgI lattice. In a low-brightness replica of the image (top right) it can be seen that the AgI structural information is nevertheless preserved. A high-pass Gaussian filtered image (bottom left) exposes the lattice under also the contaminated parts. The CNN detected atom positions of the example frame are displayed on the bottom right.}
    \label{fig: CNN_contamination}
\end{figure}

\begin{figure}[t!]
    \centering
    \includegraphics[width=\textwidth]{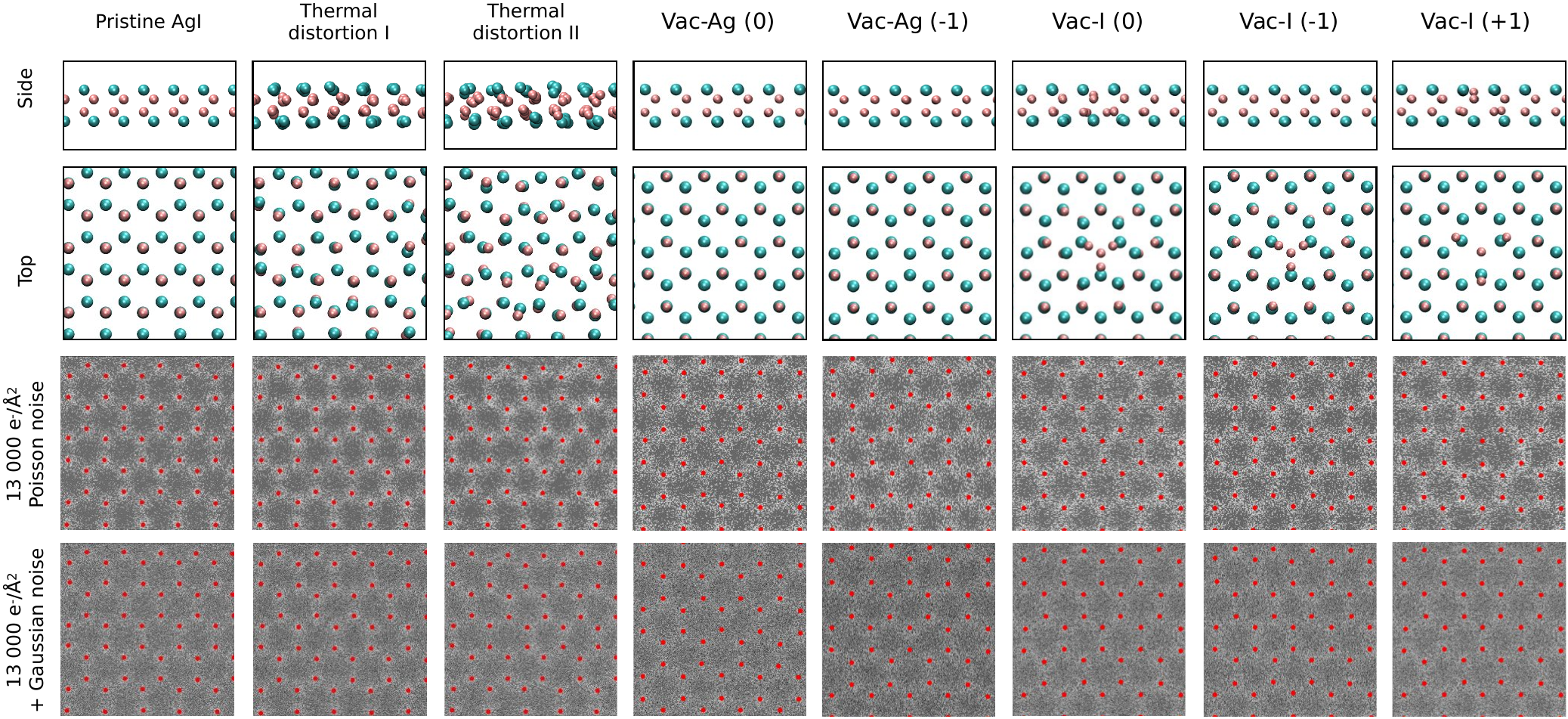}
    \caption{\textbf{Determining atom positions  in distorted AgI lattices.} The top row displays thermalized AgI structures without the $z$-directional confinement imposed experimentally by graphene, distorted by thermal excitations and vacancy-type defects with various charge states \cite{ma_defect_2023}. The middle row presents simulated ADF images computed for realistic electron doses and Poisson noise of approximately 1.3 × 10$^4$ electrons/\AA$^2$ per frame, as used in the experiments. The bottom row includes the same images with additional Gaussian noise, approximating the diffuse background from encapsulating graphene layers and the  thermal instability of the sample.}
    \label{fig:CNN_distortions}
\end{figure}

This is demonstrated for an example STEM ADF frame taken from our 1135~\degree C dataset in Figure \ref{fig: CNN_contamination}. In case the structural information is not discernible under the contamination, the CNN interprets these areas as contamination and discards them, as is explained in the discussion on the CNN architecture above.
\paragraph{CNN Handling of Vacancies and Other Lattice Distortions} The CNN used in this work was trained to detect atom positions in simple hexagonal materials. AgI, with its more complex structure comprising two atoms per column, Ag and I, poses challenges due to potential misalignment of the Ag and I atoms due to structural excitations, which could potentially confuse the CNN. Additionally, the 60 keV electron beam may in certain cases induce Ag or I vacancies, which can occasionally have nonzero charged states due to ionization \cite{ma_defect_2023}, further distorting the AgI lattice. To assess the robustness of the CNN against these lattice distortions, the CNN was tested with simulated input data shown in Figure \ref{fig:CNN_distortions}. The results indicate that the CNN is able to track the center of mass of even severely misaligned Ag and I atoms, while in case of single atom vacancies, the CNN accurately estimates the position of the remaining I and Ag atoms.

In addition to the factors discussed above, the atom positions at the image boundaries are often difficult to estimate reliably, and due to the electron-beam fly back at fast scan rates, the left edge of the acquired images is often distorted. To mitigate the effect of these uncertainties we discarded a ca. 1 nm wide slice of detected atom positions around the image perimeter from all data (see Figure \ref{fig:retrieved}a).

\section*{Comparison of Frequency Components of the CNN output and ADF Images}
\begin{figure}[!ht]
    \centering
    \includegraphics[width=0.8\textwidth]{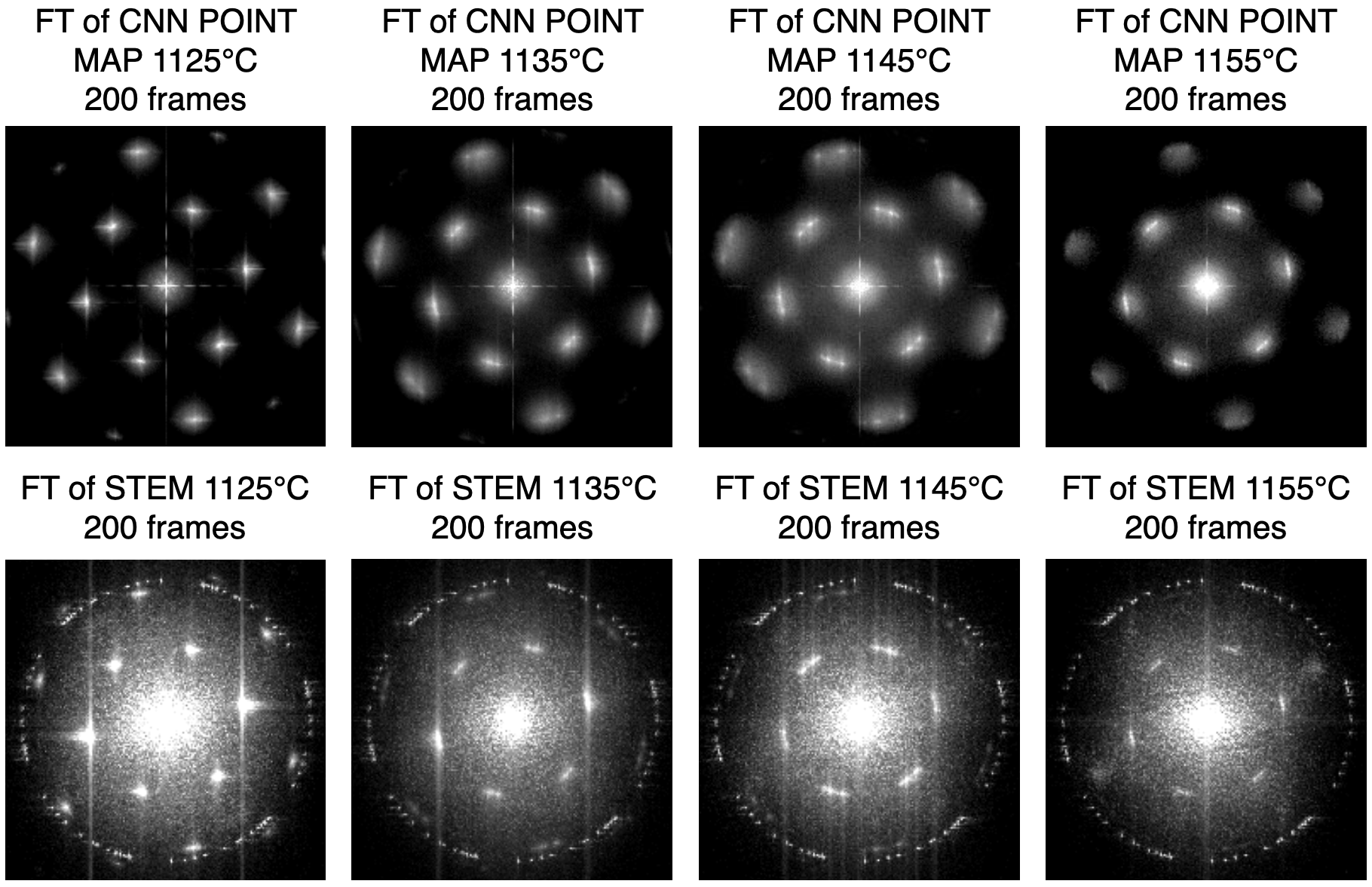}
    \caption{\textbf{Comparison of the Fourier transforms of CNN atom position maps and ADF images.} The top row shows FTs computed from CNN identified atom positions for the datasets used in spatial correlation function analysis. The bottom row presents FTs derived from the original experimental ADF images. The strong similarity between the two confirms that the atom positions identified by the CNN closely correspond to the actual atomic structure.}
    \label{fig:CNN_FT}
\end{figure}

One way to assess the reliability of the CNN output is to compare the power spectrum of the original ADF images with that of the corresponding atom position maps. If the CNN were to “hallucinate” atoms where no signal exists, these artificial atoms would introduce modes in the power spectrum that are absent in the ADF images. This comparison is shown in Figure~\ref{fig:CNN_FT}.
\section*{The Structure of 2D AgI}
\begin{figure}[ht]
    \centering
    \includegraphics[width=\textwidth]{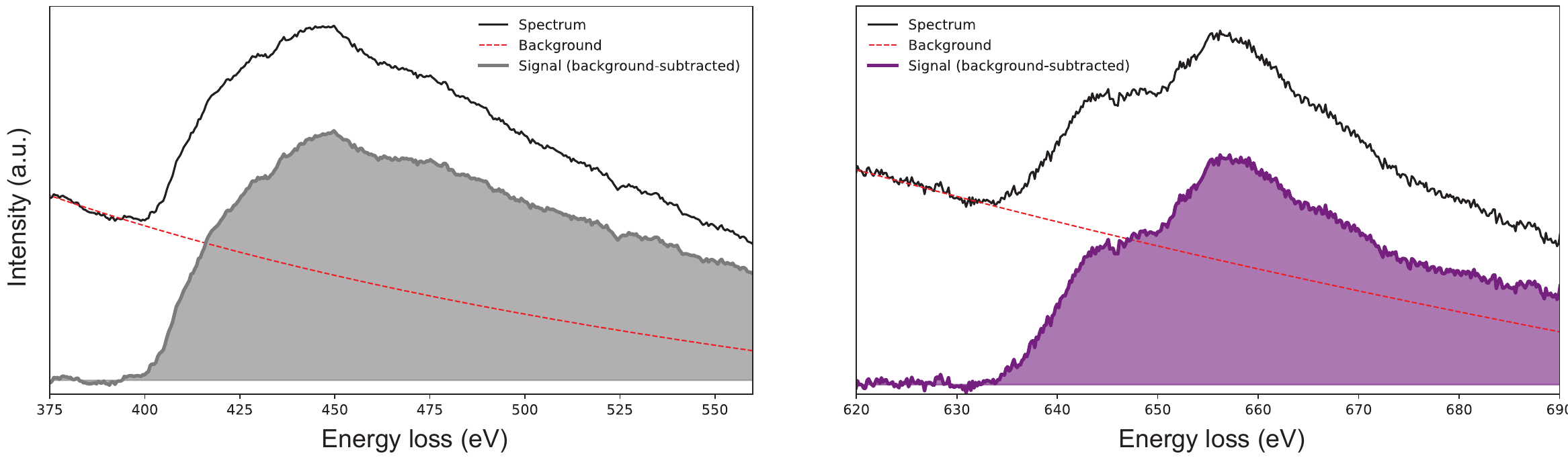}
    \caption{\textbf{Electron energy loss spectra of AgI.} The Ag M$_{4,5}$ edge is shown on the left and I~M$_{4,5}$ edge on right. The background is fitted with a power-law function.}
    \label{fig: EELS}
\end{figure}

The chemical structure of AgI was determined based on real-space imaging, electron diffraction, and EEL spectroscopy. These approaches allow us to determine an in-plane lattice parameter for AgI of 4.55 $\pm$ 0.04 \AA, which agrees well with 4.59~\AA~obtained from first principles by us and others \cite{ma_defect_2023}. The electron energy loss spectra showing the Ag and I core-loss edges is presented in Figure  \ref{fig: EELS}.

\begin{figure}[t!]
    \centering
\includegraphics[width=0.90\textwidth]{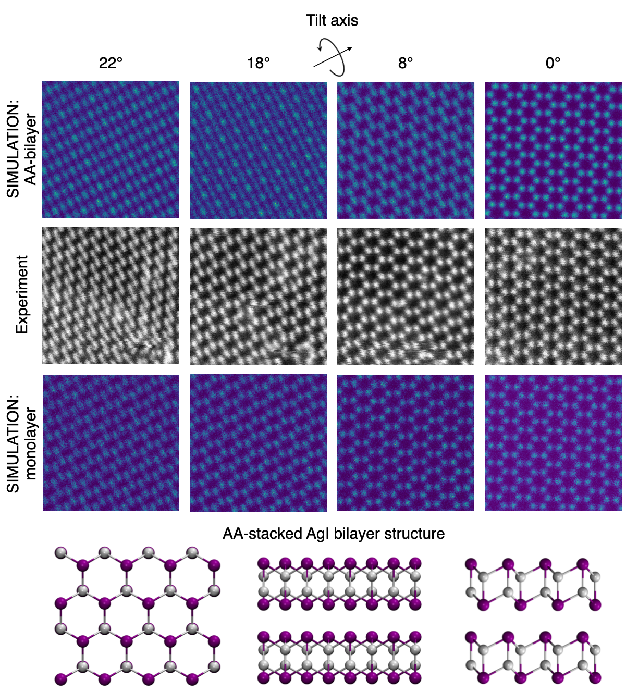}
    \caption{\textbf{Tilted STEM-ADF projections of an AgI crystal.} Comparison of experimental STEM-ADF images of an AgI crystal tilted around the indicated axis with simulated images of AA-stacked bilayer AgI (top row) and monolayer AgI (bottom row). The electron dose both in experiment and simulations was $4.2 \times 10^4$ electrons/\AA$^2$. The structure used for bilayer STEM simulations is shown below.}
    \label{fig:tilt}
\end{figure}

It is not immediately evident whether the AgI crystals in our study are monolayers, or if they, in fact, compose of two (or more) individual AgI layers in AA-stacking. In normal projection (i.e. the electron beam pinning the structure from the surface normal direction), both configurations should display exactly the same contrast, as demonstrated in Figure \ref{fig:tilt}. To resolve this, we conducted additional experiments where AgI crystals were tilted into various projections, and then the resulting ADF contrast was compared to image simulations of both AA-stacked AgI bilayers and monolayers. As can be seen from Figure \ref{fig:tilt}, the simulated monolayer contrast matches perfectly with the experiment in all accessed projections, whereas the AA-stacked bilayer yields a poor match beyond the 8\degree tilt angle, confirming the monolayer nature of the structure.

\section*{Graphene-AgI Interaction Strength}

\begin{figure}[b!]
    \centering
\includegraphics[width=0.95\textwidth]{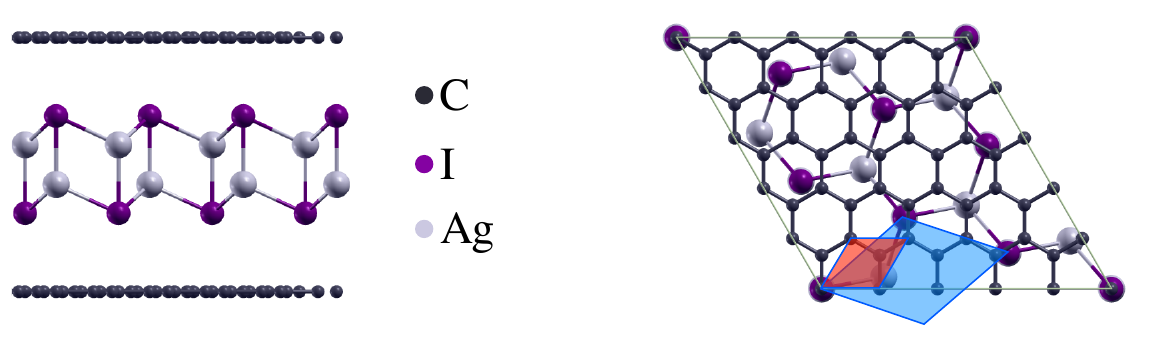}
    \caption{\textbf{Atomic structure of two-dimensional AgI encapsulated in graphene sheets.} Side (left panel) and top (right panel) view of the atomic structure with atoms indicated by colors. Red and blue shaded areas depict the primitive unit cells of graphene and AgI  lattices, respectively. The relative twist angle between the unit cells is 19.1\degree}.
    \label{fig:structure}
\end{figure}

\begin{figure}[t!]
    \centering
    \includegraphics[width=0.95\linewidth]{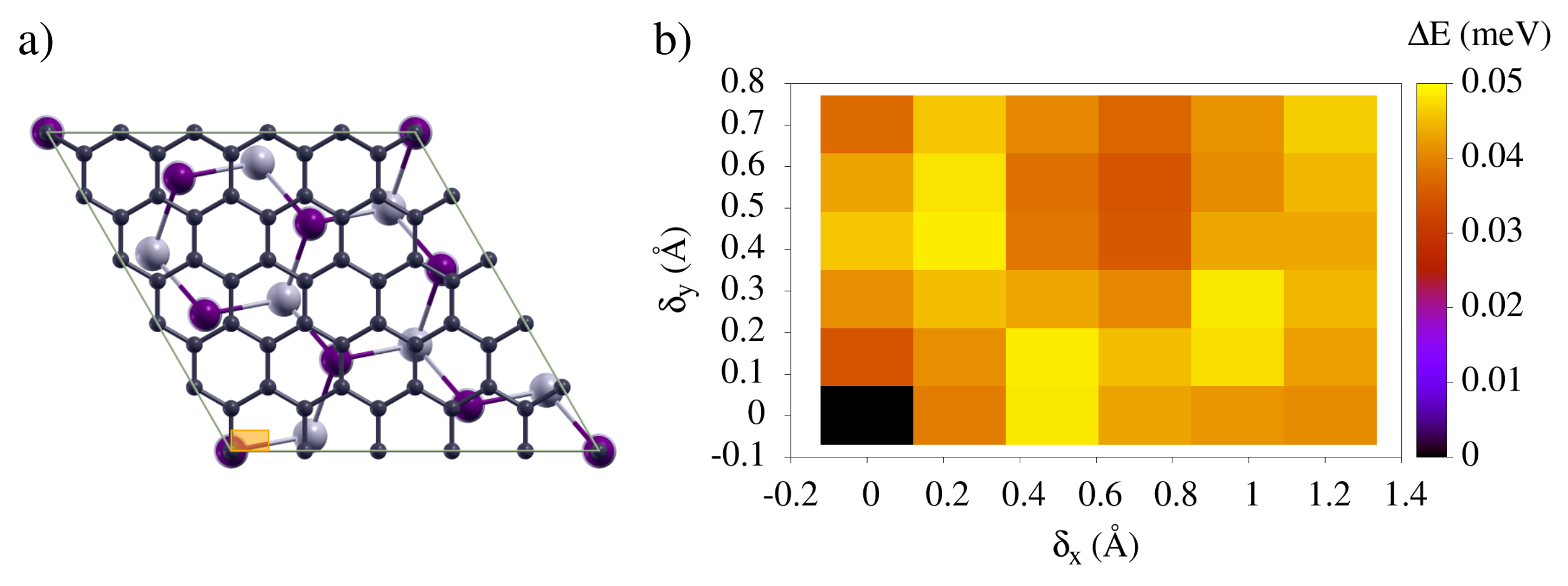}
    \caption{\textbf{Total energy profile of the G/AgI/G heterostructure under the shift of AgI layer.} \textbf{a} crystalline structure of G/AgI/G heterostructure. The orange rectangle depicts the range of shifts $(\delta_x,\delta_y)$ of AgI layer with respect to fixed graphenes. \textbf{b} Total energy difference map $\Delta E = E(\delta_x,\delta_y) - E(0,0)$ plotted for 36 positions of AgI obtained by shifting .  The $\delta_x$ and $\delta_y$ are measured from the bottom left corner of the rectangle.  $E(0,0)\equiv E(\delta_x=0,\delta_y=0)$ is the total energy corresponding to $\delta_x=0$, $\delta_y=0$, when the Iodine atom in the bottom left corner of the rectangle sits between two carbon atoms . }
    \label{fig:en_map}
\end{figure}

Due to the difference in lattice constants, the graphene and AgI lattices are inherently non-commensurate. To estimate the interaction strength between these structures from first principles, we utilized the CellMatch code~\cite{LAZIC2015324} to identify feasible commensurability conditions. This approach enabled the construction of a relatively small supercell containing only 128 atoms, where the AgI layer was sandwiched between graphene layers. In this configuration, the graphene lattice constant was compressed by 1.2\% to $a_{\text{G}} = 2.43$~\AA, and the AgI layer was twisted by $\theta$ = 19.1\degree~relative to the graphene layers; without the twist and compression the simulation cell would be too large to compute the interaction energies. The final structure is shown in Figure \ref{fig:structure}.
Within the simulation cell, it is evident that the positions of iodine (I) atoms closest to the graphene layers exhibit no strong correlation with the carbon atoms. Instead, these iodine atoms are distributed almost equally among the top sites (I atom positioned directly below a carbon atom), bridge sites (I atom located at the midpoint of a C–C bond), and hollow sites (I atom situated at the center of a graphene hexagon). It is therefore anticipated that AgI will exhibit only a weak positional preference relative to the graphene layers. This is confirmed by performing total energy calculations for 36 distinct positions of AgI while keeping the graphene layers fixed. Each configuration is made by shifting the AgI atoms from their initial position shown in Figure \ref{fig:en_map}a by $\delta_x \in [0,1.2145~\mathring{A}]$,  $\delta_y \in [0,0.7012~\mathring{A}]$. Due to symmetry of graphene lattice we change the position of AgI by $(\delta_x$,$\delta_y)$ in the range depicted by the orange rectangle in Figure \ref{fig:en_map}a. For all configurations the interlayer distance is fixed at $d_{\text{inter}}=3.42$~\AA, which corresponds to the average interlayer distance of the fully relaxed structure for $\delta_x=0$ and $\delta_y=0$. 
The obtained total energy profile $\Delta E = E(\delta_x,\delta_y) - E(0,0)$ shown in Figure \ref{fig:en_map}b is nearly uniform, indicating that the energy landscape is flat, with a variation of only 1 meV across the entire simulation cell (128 atoms). Considering that the commensurability of the graphene-AgI system in real samples is lower than in our simulations (where compressed graphene ensures commensurability within a cell of approximately 1 nm), it is expected that, in actual samples where graphene layers are randomly oriented and non-compressed, the energy landscape will be even flatter. This directly implies that the AgI crystal is nearly free of orientational constraints, and that the mutual orientation and alignment between AgI and graphene is not important for the melting process.

\section*{Image Artifacts Related to Iodine Impurities}
Chains of bright spots are frequently observed both on AgI crystals and in the surrounding areas during our STEM experiments. These features typically become more prominent as the temperature increases; this is for instance the case in Figure~\ref{fig:2D_melting} (and in Figures \ref{fig:2D_melting_extras_1}-\ref{fig:2D_melting_extras_4}, where temperatures in excess of 1000~\degree C are accessed.

We hypothesize that these bright spots and ribbons correspond to iodine atoms trapped at the edges of contamination that becomes graphitized at sufficiently high temperatures. At room temperature, such chains are rarely seen, although isolated bright atoms dispersed within the amorphous material on graphene are commonly present.

\begin{figure}
    \centering
    \includegraphics[width=0.9\linewidth]{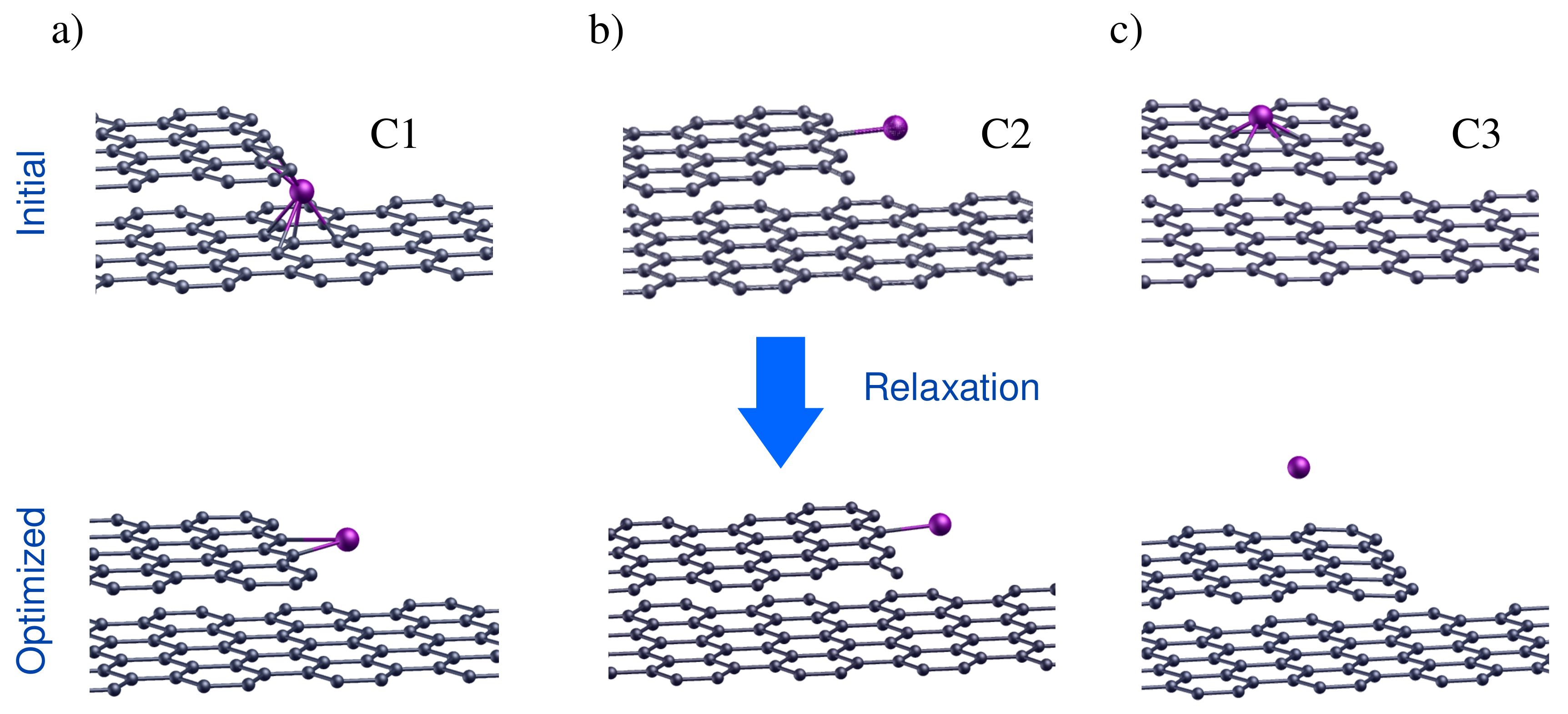}
    \caption{\textbf{Adsorption of an iodine atom near the edge of a graphene pore.} Initial atomic configurations are shown in the top row. After the structure optimization the I atom moved towards the edge of the upper carbon layer (a), (b) or was repelled from graphene surface increasing the distance from graphene and total energy (c).  The binding (adsorption) energy  $E_{\rm{b}}=E_{\rm{I,G}}-E_{\rm{G}}-E_{\rm{I}}$ is  $E_{\rm{b}}^1=-3.771$\,eV,  $E_{\rm{b}}^2=-4.25$\,eV and  $E_{\rm{b}}^3=-0.879$\,eV for C1, C2 and C3 configuration, respectively, where $E_{\rm{I,G}}$ is the total energy of the full (graphene + iodine) system, $E_{\rm{G}}$  is the the total energy of graphene without iodine and $E_{\rm{I}}$ is the total energy of an isolated iodine atom.} 
    \label{fig:I_adsorption}
\end{figure}

\begin{figure}
    \centering
    \includegraphics[width=0.6\linewidth]{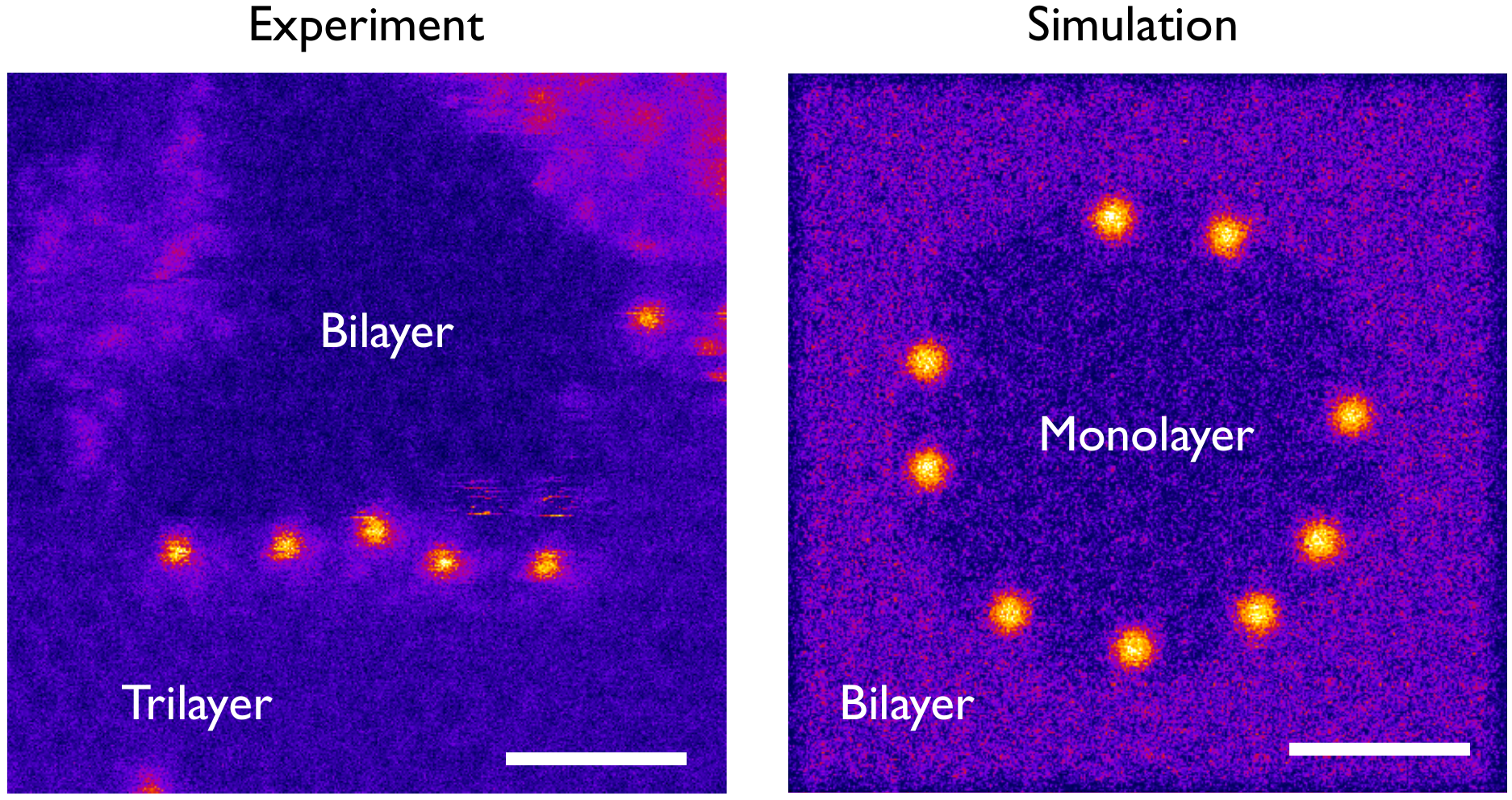}
    \caption{\textbf{STEM ADF images and simulations of iodine atoms at graphene pores.} The left panel shows an experimental STEM-ADF image of iodine atoms bound at the edge of a graphene pore. The thinner region is a bilayer, with iodine atoms bound at the edge of a third layer. The right panel shows a simulated STEM-ADF image, where the thinner region is a monolayer and the iodine atoms are attached at the edge of a second layer. The scale bars are 2 nm.} 
    \label{fig:iodine_STEM}
\end{figure}

To evaluate this hypothesis, we modeled a pore in bilayer graphene by placing a graphene nanoribbon on an infinite graphene sheet. An iodine atom was positioned near the pore edge, and the structure was relaxed, allowing the iodine and nanoribbon atoms to move while keeping the bottom-layer carbon atoms fixed. The resulting relaxed structures are shown in Figure~\ref{fig:I_adsorption}. The binding energies for the C1 and C2 configurations are –3.771~eV and –4.250~eV, respectively—both significantly lower than that of the C3 configuration (–0.879~eV). These results indicate that iodine atoms preferentially bind to graphene edges, where they can minimize energy by saturating the dangling bonds of carbon atoms.
\newpage

\section*{2D to 3D Transformation of AgI Upon Heating}

\begin{figure}[ht!]
    \centering
    \includegraphics[width=1.0\textwidth]{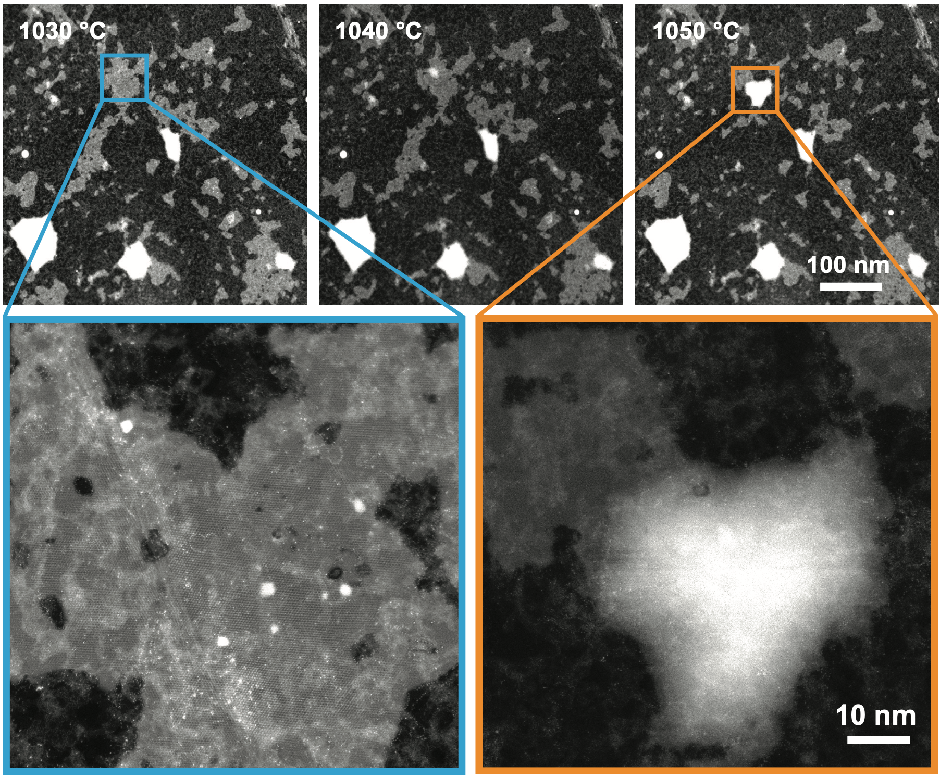}
    \caption{\textbf{AgI crystals undergoing 2D to 3D transformation upon heating.} The top row shows a group of graphene-encapsulated AgI crystals at 1030~\degree C. At 1040~\degree C (middle), a small 3D pocket of supercritical AgI fluid forms, which grows in size and engulfs most of the crystal by 1050~\degree C.} 
    \label{fig:2D-3D-transformation}
\end{figure}
\newpage
\section*{Supplementary Datapoints for Manuscript Figure \ref{fig:2D_melting}}
\begin{figure}[bh!]
    \centering
    \includegraphics[width=12.5cm]{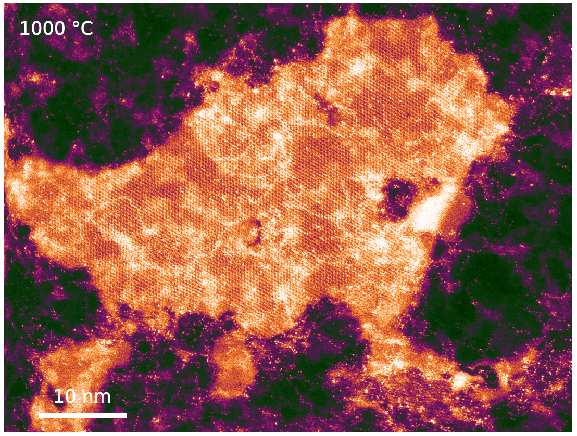}
    \caption{\textbf{1000~\degree C.}}
    \label{fig:2D_melting_extras_1}
\end{figure}
\newpage
\begin{figure}[th!]
    \centering
    \includegraphics[width=12.5cm]{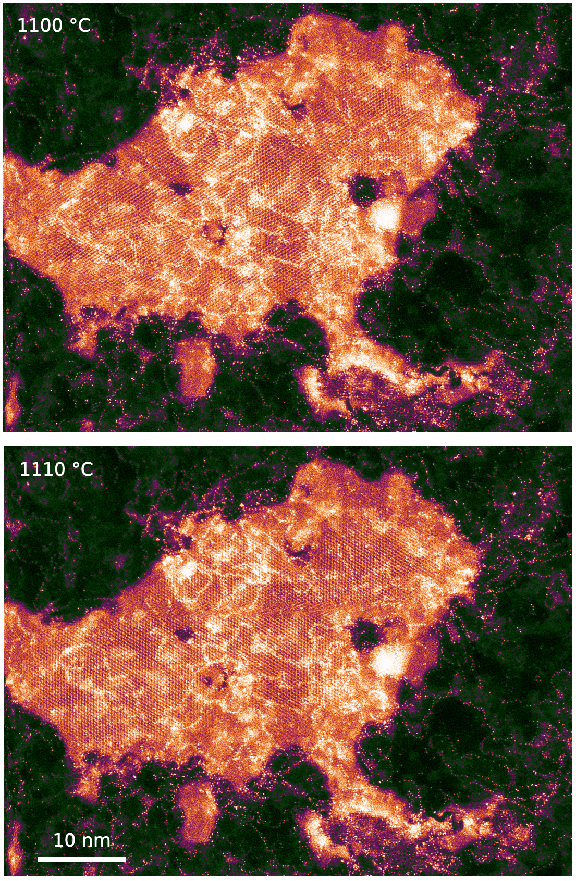}
    \caption{\textbf{1100~\degree C and 1110~\degree C.}}
    \label{fig:2D_melting_extras_2}
\end{figure}
\begin{figure}[th!]
    \centering
    \includegraphics[width=12.5cm]{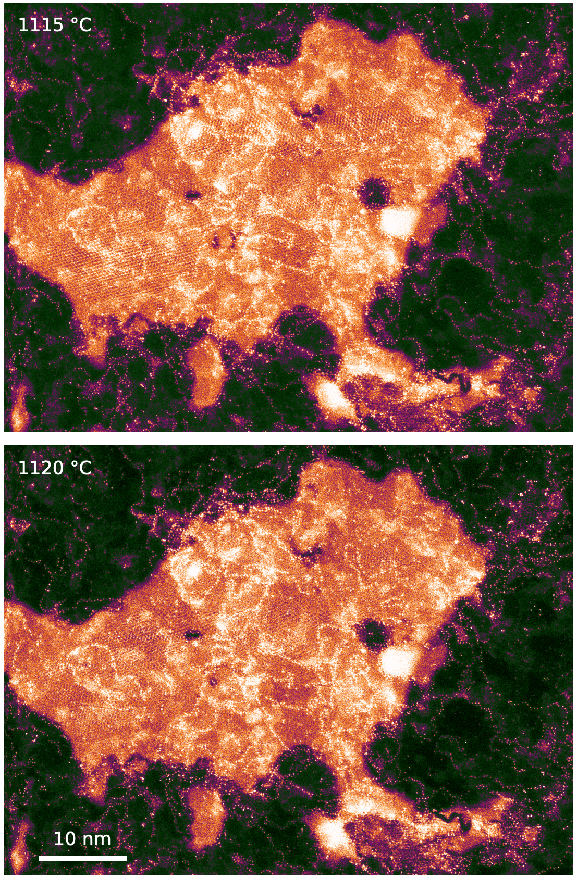}
    \caption{\textbf{1115~\degree C and 1120~\degree C.}}
    \label{fig:2D_melting_extras_3}
\end{figure}
\begin{figure}[th!]
    \centering
    \includegraphics[width=12.5cm]{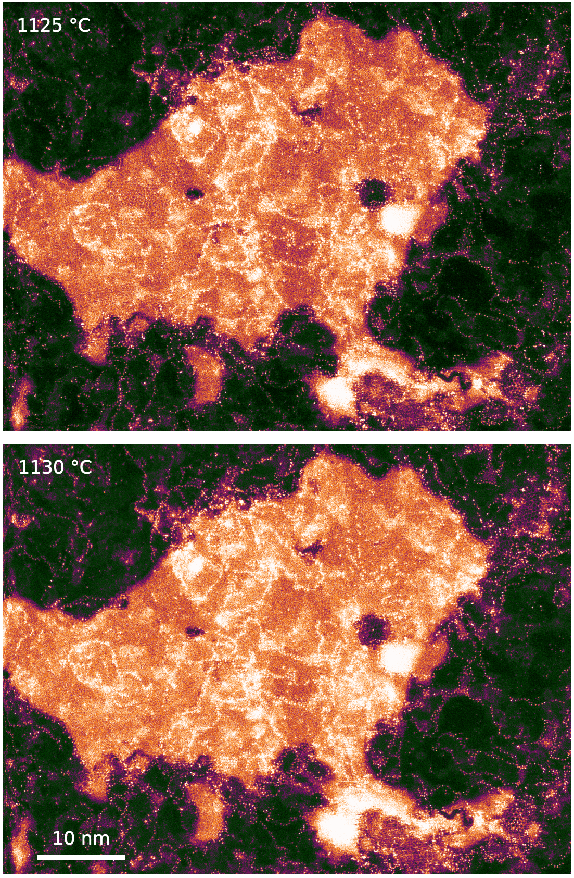}
    \caption{\textbf{1125~\degree C and 1130~\degree C.}}
    \label{fig:2D_melting_extras_4}
\end{figure}
\newpage
\section*{Small AgI Crystal Undergoing Phase Transitions}

\begin{figure}[ht!]
    \centering
    \includegraphics[width=0.9\textwidth]{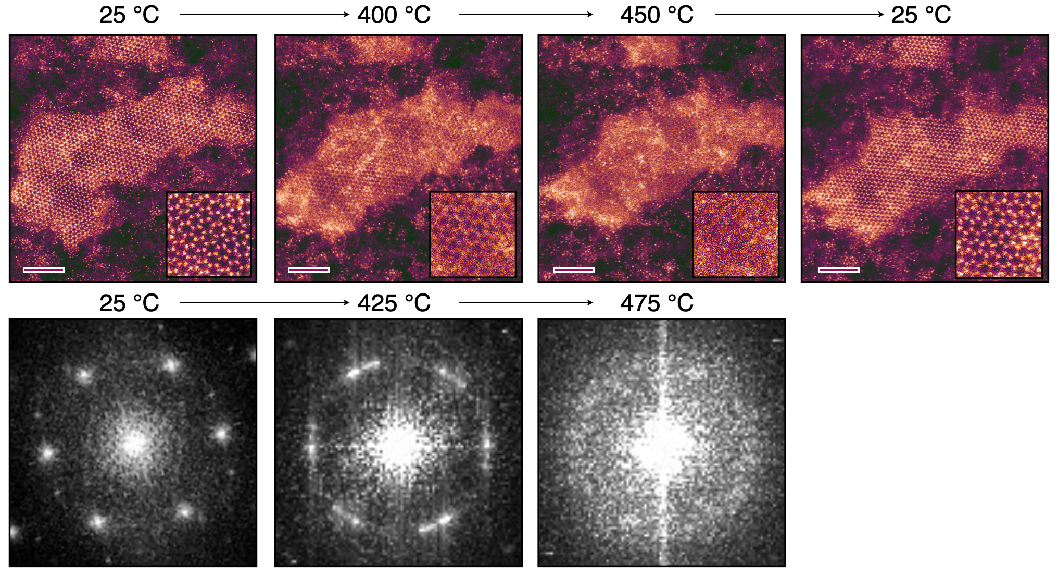}
    \caption{\textbf{A small AgI crystal going through the transitions from solid to hexatic to liquid phase.} The FTs on the lower row are composition of 100 rapidly acquired ADF images. Note that the high temperature FT data were acquired at temperatures 25~\degree C higher than the real-space ADF images. The scale bars are all 2 nm.}
    \label{fig: AgI_small}
\end{figure}
\newpage
\section*{2D-Melting-Point Depression}

\begin{figure}[t!]
    \centering
    \includegraphics[width=0.9\textwidth]{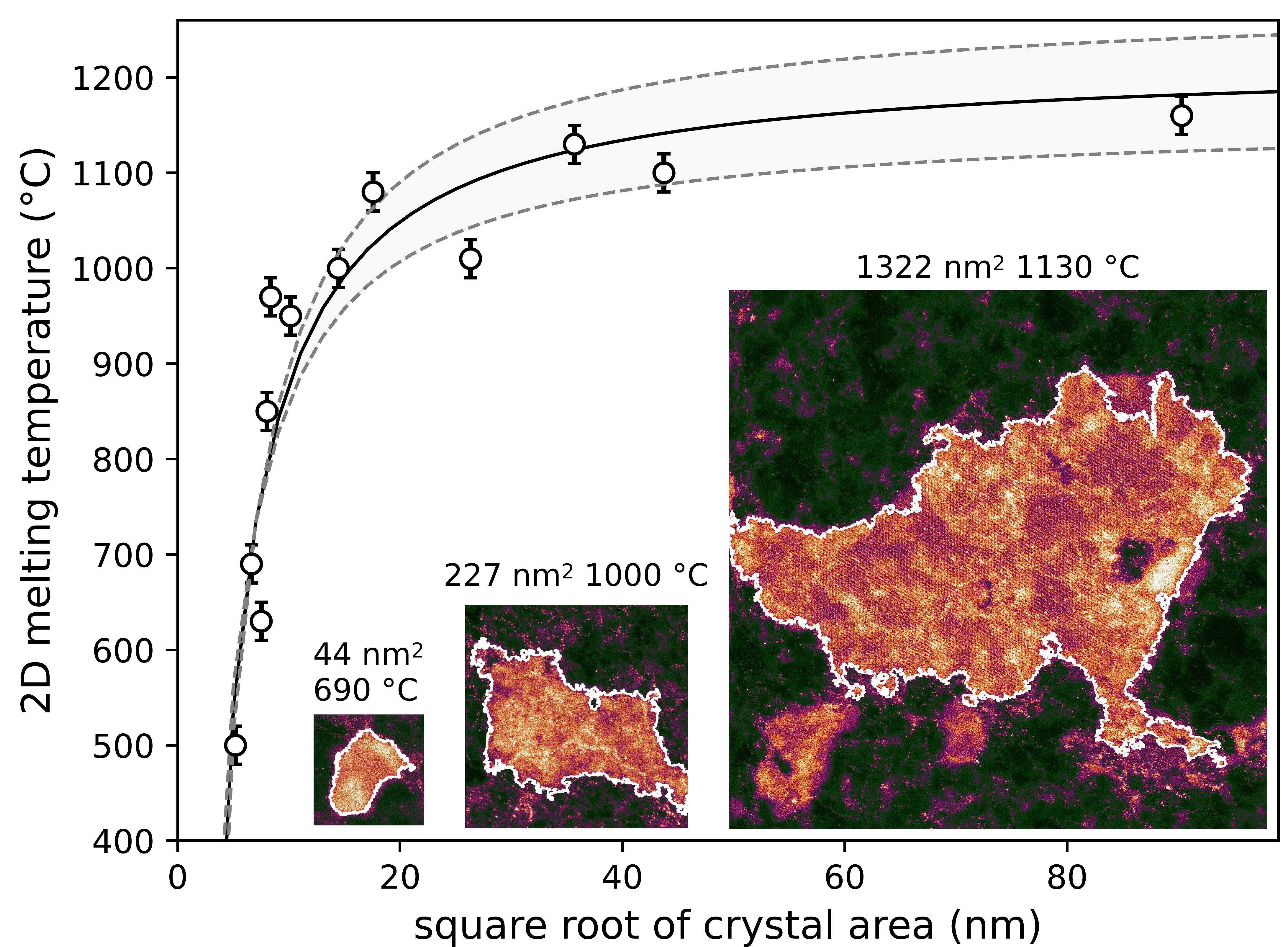}
    \caption{\textbf{Melting-point depression in 2D.} The insets show example STEM ADF images of crystals studied for the graph, and they are all displayed in the same scale. The solid line is the best fit to \ref{equ: melting_temp} and the dashed lines represent 95\% confidence bounds of the fit.}
    \label{fig:temperature}
\end{figure}

At the nanometer scale, it is well established that the melting temperature of solids exhibits size dependence. This effect arises from differences in cohesive energy between surface and bulk atoms. For 3D nanoparticles, the melting temperature is inversely proportional to the square of the particle diameter or, equivalently, to the surface-to-volume ratio \cite{buffat_size_1976}. In 2D particles, this scaling changes, reflecting the ratio of the circumference to the area, resulting in an inverse proportionality to the particle diameter~\cite{zhang_size_2018}. Figure \ref{fig:temperature} illustrates this relationship for various 2D AgI crystal sizes, with the $x$-axis representing the square root of the particle surface area. The surface area was determined from calibrated STEM ADF images using the ImageJ software suite. The experimental data in Figure \ref{fig:temperature} is fitted to the model
\begin{equation}
  T_{m,2D} = T_{m,2D\infty} \left(1 - \frac{B}{\sqrt{A}} \right),
    \tag{Eq. S2}\label{equ: melting_temp}
\end{equation}
where \( B \) and \( T_{m,2D-\infty} \)~are fitting parameters, and $A$ is the area of the 2D particle. Based on this model, we estimate the 2D melting point for a crystal of infinite size, \( T_{m,2D\infty} \), to be 1220$\pm$40~\degree C, about 600~\degree C higher than the reported bulk value. This large increase is expected in van der Waals encapsulation, where the graphene layers provide strong confinement and generally enhance the thermal stability of encapsulated solids compared to the free-standing bulk.

\clearpage

\section*{Broadening of Diffraction Profiles in the Liquid Phase}
\begin{figure}[!ht]
    \centering
    \includegraphics[width=\textwidth]{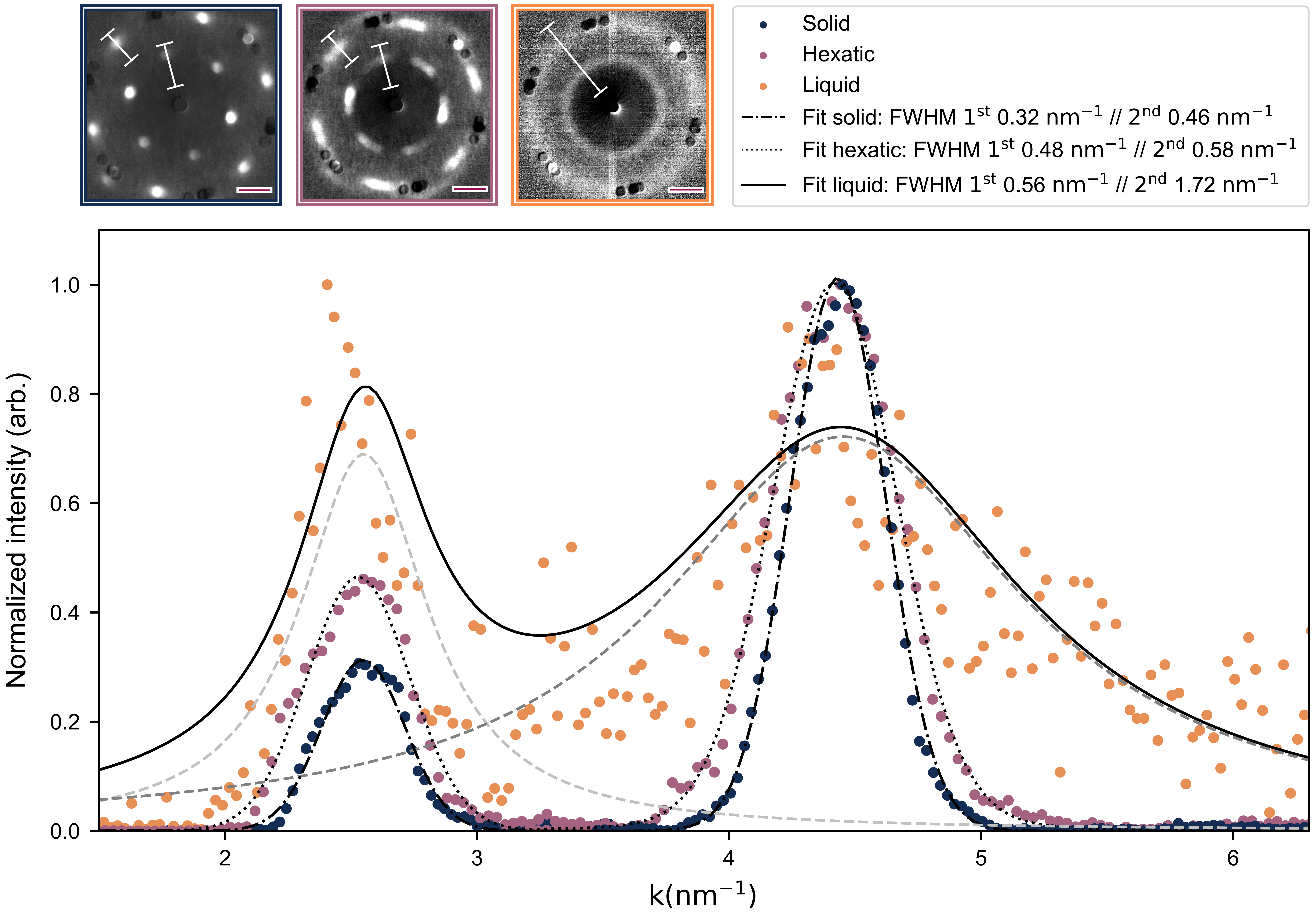}
    \caption{\textbf{Broadening of the diffraction features} Line profiles over the diffraction features in Figure~\ref{fig:NBED} are analyzed to extract the full width at half maximum (FWHM) of each diffraction feature. The data is fitted using a sum of Gaussian and Cauchy-Lorentz line shapes (Voigt profile).}
    \label{fig:line_profiles}
\end{figure}

\begin{figure}[!ht]
    \centering
    \includegraphics[width=\textwidth]{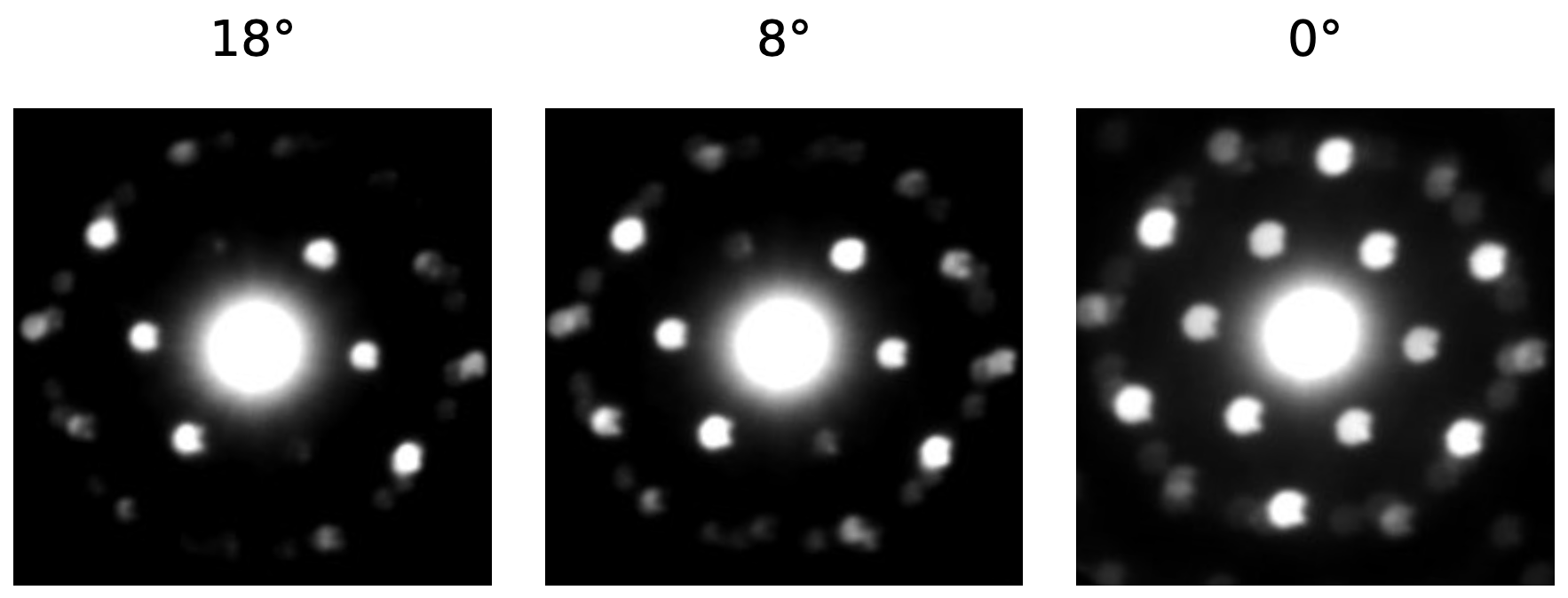}
    \caption{\textbf{Diffraction patterns of a tilted AgI crystal.} The variation in diffraction intensity observed in manuscript Figure~\ref{fig:NBED} can be attributed to unevenness in the encapsulating graphene and to local tilts across different regions of the AgI crystal. This effect is demonstrated here by intentionally tilting a flat AgI crystal to various angles and acquiring the corresponding diffraction patterns.}
    \label{fig:diffraction_tilted}
\end{figure}

Figure \ref{fig:line_profiles} shows averaged diffraction patterns and line profiles from the solid, hexatic and liquid phase. It is assumed that the FWHM measured in the solid phase is dominated by the intrinsic width as given by the NBED convergence angle, and additional broadening in the liquid and hexatic phase is added in quadrature. Hence, the broadening due to disorder in the liquid phase is estimated as 
\begin{equation}
\Delta k_{liquid}=\sqrt{(\Delta k_{total})^2-(\Delta k_{solid})^2},
    \label{equ: correlation_length}
    \tag{Eq. S3}
\end{equation}
which results in 1.66~nm$^{-1}$ for the second order ring and $0.46~\textrm{nm}^{-1}$ for the first order ring. This means that the distance ($1/\Delta k$) over which a periodicity and hence a correlation is present is ca. 0.6~nm for the periodicity of the second order peak (0.22~nm) and ca. 2~nm for that of the first order peak (0.38~nm). 

\clearpage
\section*{Diffraction Patterns of Tilted AgI Crystals}

\clearpage

\section*{Local Density analysis}

As different phases during the melting process commonly have different structural densities, the temporal or spatial presence of one or more phases can be determined by looking at the local density distribution of the image stacks. The local density $\rho_l$ around a point at position $r_l$ is defined as
\begin{equation}
\tag{Eq. S4}
    \rho_l= \sum_j H(r_c -|r_l-r_j|/\pi r_c^2),
\end{equation}
\noindent where $H$ is the Heaviside step function and $r_c$ is the range around which the density is analyzed. To obtain meaningful results, $r_c$ should be significantly smaller than the total system size, yet substantially larger than the lattice constant $a_0$. In our systems we obtain robust results for $3a_0<r_c<6a_0$, and therefore chose $r_c=5a_0$ for this analysis.

As the segmentation step of the CNN analysis introduces "holes" in the atomic position files, which results in incorrect datapoints in the lower range of the local density distribution, we excluded from the analysis all points whose local density radius $r_c$ includes at least one position adjacent to such a hole. 

We applied the following criterion to identify points bordering a "hole": if a point has fewer than five neighbors within a radius $r_s$ = 1.2$a_0$, it is considered adjacent to a segmentation-induced hole and excluded from the analysis. The value of $r_s$ was selected to effectively remove the most significant artifacts in the lower end of the local density distribution, while still retaining sufficient data at the highest temperatures, where the local density is naturally lower. Despite this filtering, the broad spread in local density observed at 1155~\degree C and 1160~\degree C remains partially influenced by the holes introduced during the segmentation step.

Up to 1125~\degree C, we observe very little change in the local density distribution averaged over all frames, as expected for a fully solid material. Starting at 1125~\degree C, the distribution begins to broaden gradually and shifts slightly toward lower densities. At 1135~\degree C, a distinct lower-density mode emerges, which grows more prominent with increasing temperature and becomes dominant by 1150~\degree C. In this temperature range, the density distribution exhibits a bimodal character, as expected for a first-order transition. We also observe strong frame-to-frame fluctuations.

At 1155~\degree C, we observe a near-complete collapse of local order, and the density distribution becomes significantly broadened. It is important to note that this analysis is limited by the relatively small size of the acquired STEM images, especially when compared to the large number of particles typically accessible in simulations. Temporal fluctuations are observed as strong variations in local density between consecutive frames. Similarly, spatial coexistence of well-ordered and disordered regions is evident within individual frames. However, due to the limited field of view, we were unable to extract a reliable estimate of the typical domain size.

\clearpage
\section*{Defect analysis}

To analyze how defect structures evolve with increasing temperature, we identified topological defects using Voronoi tessellation applied to the set of polygon centers. Each point was assigned a Voronoi cell, and any non-hexagonal cell was classified as a defect. 

We first plot the evolution of 5–7 pairs, isolated dislocations (5–7 pairs entirely surrounded by hexagons), and isolated disclinations (individual 5- or 7-coordinated cells). Below 1120~\degree C, the density of these defects remains low and is dominated by vacancy-like defects—visualized in the Voronoi diagrams as higher-order polygons surrounded by two or three 5-sided cells—as well as 5–7–7–5 defects. The majority of these structures carry zero net Burgers vector and therefore preserve long-range translational and orientational order. 

While some images show isolated dislocations, detailed analysis reveals that most 5–7 pairs align into lines with a net zero Burgers vector (Figure~\ref{fig: burgers}). As the temperature approaches 1125~\degree C, the overall defect density increases steadily, signifying the onset of the hexatic phase. Between 1125~\degree C and 1140~\degree C, the density of isolated dislocations rises sharply. Above 1140~\degree C, however, 5–7 pairs increasingly aggregate into extended defect clusters, leading to a decline in the number of isolated dislocations. 

At lower temperatures, free disclinations are rare. However, beginning around 1130~\degree C, isolated 5- and 7-fold coordinated defects — interpreted as disclination-like quasi-particles — start to appear. The observed increase in defect density and the unbinding of dislocation pairs near 1125~\degree C are consistent with the KTHNY theory, which predicts dislocation unbinding at the solid-to-hexatic transition. Although direct evidence of dislocation unbinding into disclinations at the hexatic-to-liquid transition remains elusive, the steady rise in isolated disclinations at these temperatures implies a continuous Kosterlitz-Thouless-type transition.

Since the number of strictly isolated topological defects in our system is very low compared to the overall defect density, it is evident that most defects aggregate into larger clusters even at relatively moderate temperatures. This clustering introduces significant ambiguity in the identification and classification of individual topological defects.

To address this ambiguity, we analyze the topological charge of defect clusters by computing the total disclination charge $q=\sum q_a$, where $q_a=n_a-6$ and $n_a$ is the number of edges of the Voronoi cell associated with each defect in the cluster. Clusters with $q \neq 0$ carry a net disclination charge, while those with $q = 0$ may be either neutral or dislocation-charged, depending on their net Burgers vector. The Burgers vector $\Vec{b}$ of a cluster is calculated from the positions $\Vec{r}_a$ and disclination charges $q_a$ of all defects in the cluster using the following expression:

\begin{equation}
    \Vec{b} = \Vec{z} \times \sum_a q_a \Vec{r}_a
\end{equation}

\noindent where $\Vec{z}$ is the out-of-plane unit vector. Since the Burgers vector must correspond to a lattice vector, the calculated vector was snapped to the nearest lattice point of the ideal hexagonal AgI lattice. Defect clusters with $\Vec{b} = 0$ are classified as neutral, whereas those with $\Vec{b} \neq 0$ and $q \neq 0$ are considered dislocation charged. While this method reliably characterizes defect clusters of limited size that are fully contained within the acquired frame, it becomes unreliable once clusters form extended lines that reach the frame boundaries, where Voronoi tessellation is no longer suitable for identifying defects. 

To eliminate potential ambiguities arising near the image boundaries, we implemented two measures: (1) all Voronoi cells associated with points located within a distance $a_0$ of the frame edge were artificially converted into hexagons; and (2) any defect cluster containing at least one point within $2a_0$ of the border was assigned to an “Unknown” category.

The segmentation step in the CNN analysis occasionally results in "holes" in the atomic position files, which in turn leads to "flower"-like structures with unusually elongated cells in the Voronoi tessellation. To exclude clusters containing such points from the defect analysis, we calculate the isoperimetric quotient $IQ= 4\pi A/ C^2$ of each Voronoi cell, where where $A$ is the area and $C$ is the circumference of the cell.

Regular polygons with $n$ edges have an $IQ$ parameter of $\pi /[n \times $tan$( \pi/n)]$, which approximates to 0.86, 0.91 or 0.93 for pentagons, hexagons and heptagons, respectively. Polygons with lower $IQ$ values are more irregular and elongated. In our system, most polygons exhibit slightly irregular shapes, with $IQ$ parameters ranging between 0.75 and 0.92, while the elongated polygons around the holes have $IQ$ values below 0.7. By excluding all clusters containing at least one point with an $IQ$ below this cutoff, we can obtain more accurate statistics without discarding too much data.

\begin{figure}[htp!]
    \centering
    \includegraphics[width=1\textwidth]{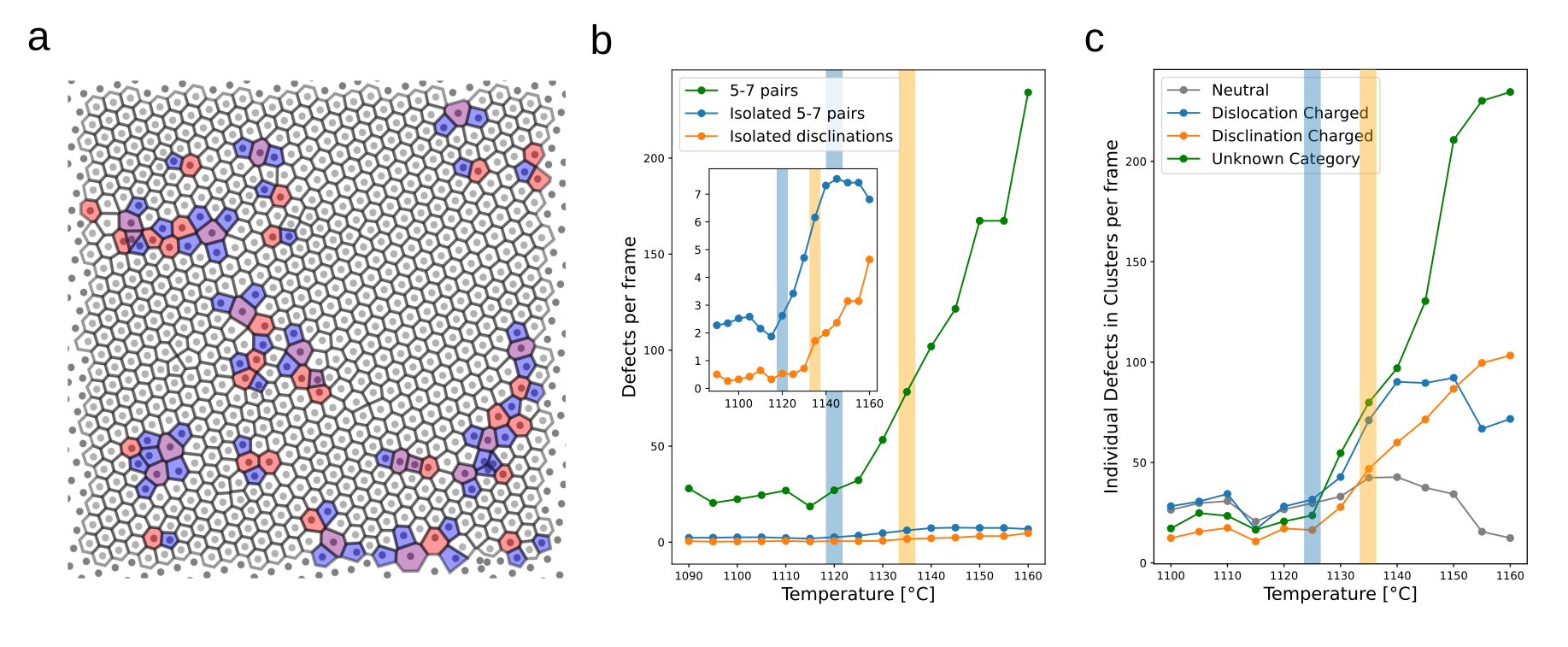}
    \caption{\textbf{Defect analysis.} \textbf{a} Cutout of a Voronoi diagram of a single frame at 1135~\degree C. Pentagons, heptagons and other not 6-sided polygons are colored in blue, red and purple, respectively. \textbf{b} The total amount of 5-7 pairs, isolated 5-7 pairs and isolated disclinations as a function of temperature. Note that the total amount of 5-7 pairs is scaled down by a factor of 10. \textbf{c} The total amount of defects contained within clusters with neutral, dislocation and disclination charge as well as within clusters that could not be categorized as a function of temperature.} 
    \label{fig: Defects}.
\end{figure}

\begin{figure}[ht]
    \centering
    \includegraphics[width=0.6\textwidth]{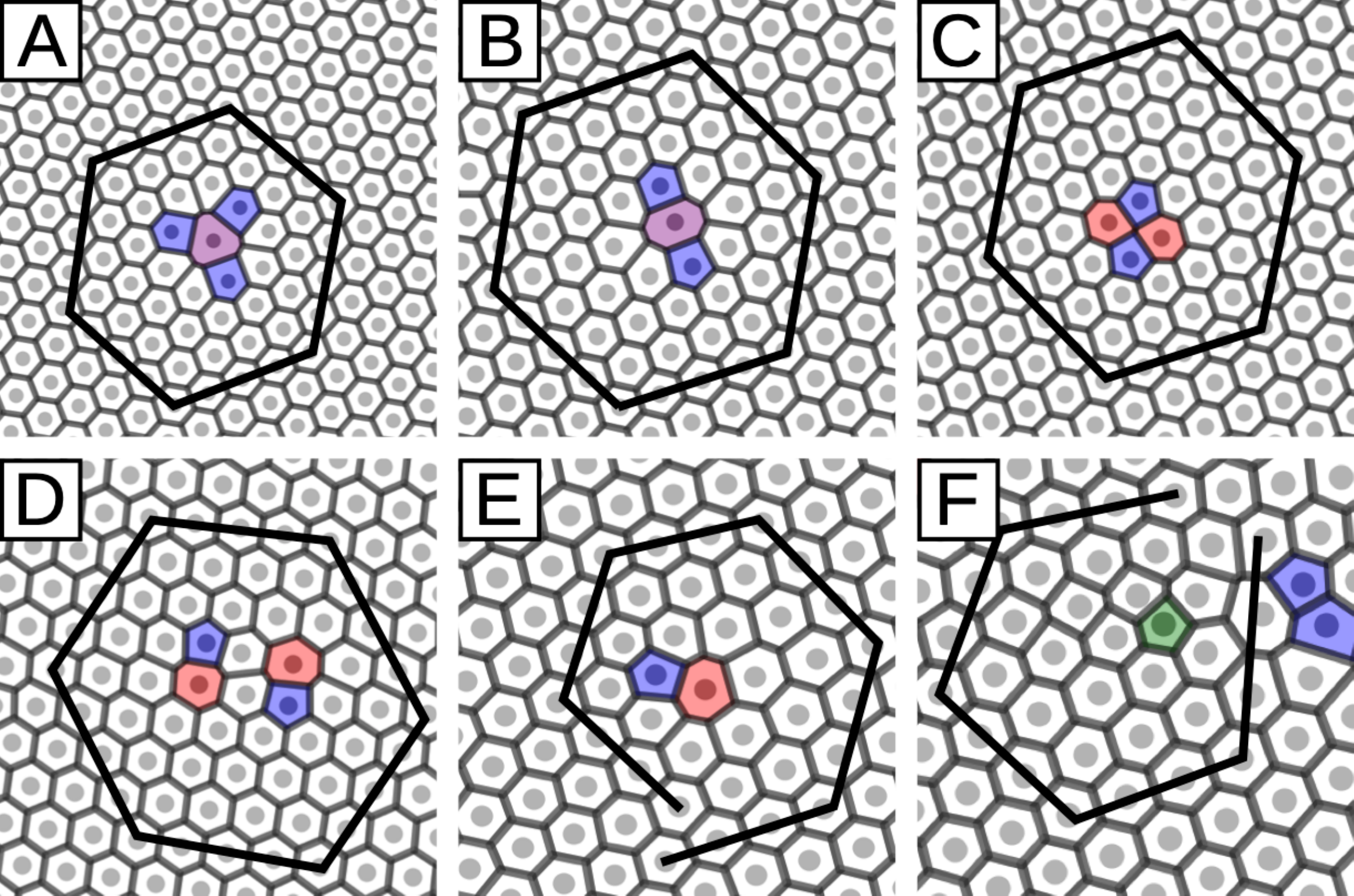}
    \caption{\textbf{Burgers vectors of defects.} Vacancy-type defects \textbf{a-b}, as well as 5-7-7-5 \textbf{c} and paired 5-7 defects \textbf{d} have a zero Burgers vector, thus preserving long range order. Dislocations (isolated 5-7 defects) (E) and disclinations (F) have a non-zero Burgers vector which leads to decay of the translational and orientational order.}
    \label{fig: burgers}
\end{figure}

Figure \ref{fig: Defects}c shows the statistics of topological defects within clusters, categorized as topologically neutral, disclination- and dislocation-charged, as well as clusters that could not be categorized due to the limited size of our system. Even at low temperatures, a stable number of topological defects is present, with the majority being neutral or dislocation-charged. It is important to note that dislocation-charged defect clusters do not necessarily disrupt long-range translational order, as they often cancel each other out, similar to paired 5-7 defects with a zero Burgers vector. Starting at 1125~\degree C, the abundance of non-neutral defect types increases rapidly, and defect clusters agglomerate more significantly, ultimately leading to a reduction in identifiable dislocation-charged clusters by 1140~\degree C. At higher temperatures, most defects form large clusters or chains, which frequently act as grain boundaries. In contrast to the analysis of strictly isolated topological defects, this analysis reveals no distinct behavior of disclination- and dislocation-charged defects, as predicted by the Kosterlitz-Thouless theory.

Finally, it should be noted that the statistics presented in Figure \ref{fig: Defects} are estimates, primarily useful for qualitative analysis. A more rigorous quantitative analysis would require a significantly larger system, which are not accessible with the present experimental techniques.

\newpage

\newpage
\section*{Burgers Vectors of Defects}
\newpage
\section*{Structure Factors S(q)}
\begin{figure}[ht]
    \centering
    \includegraphics[width=14cm]{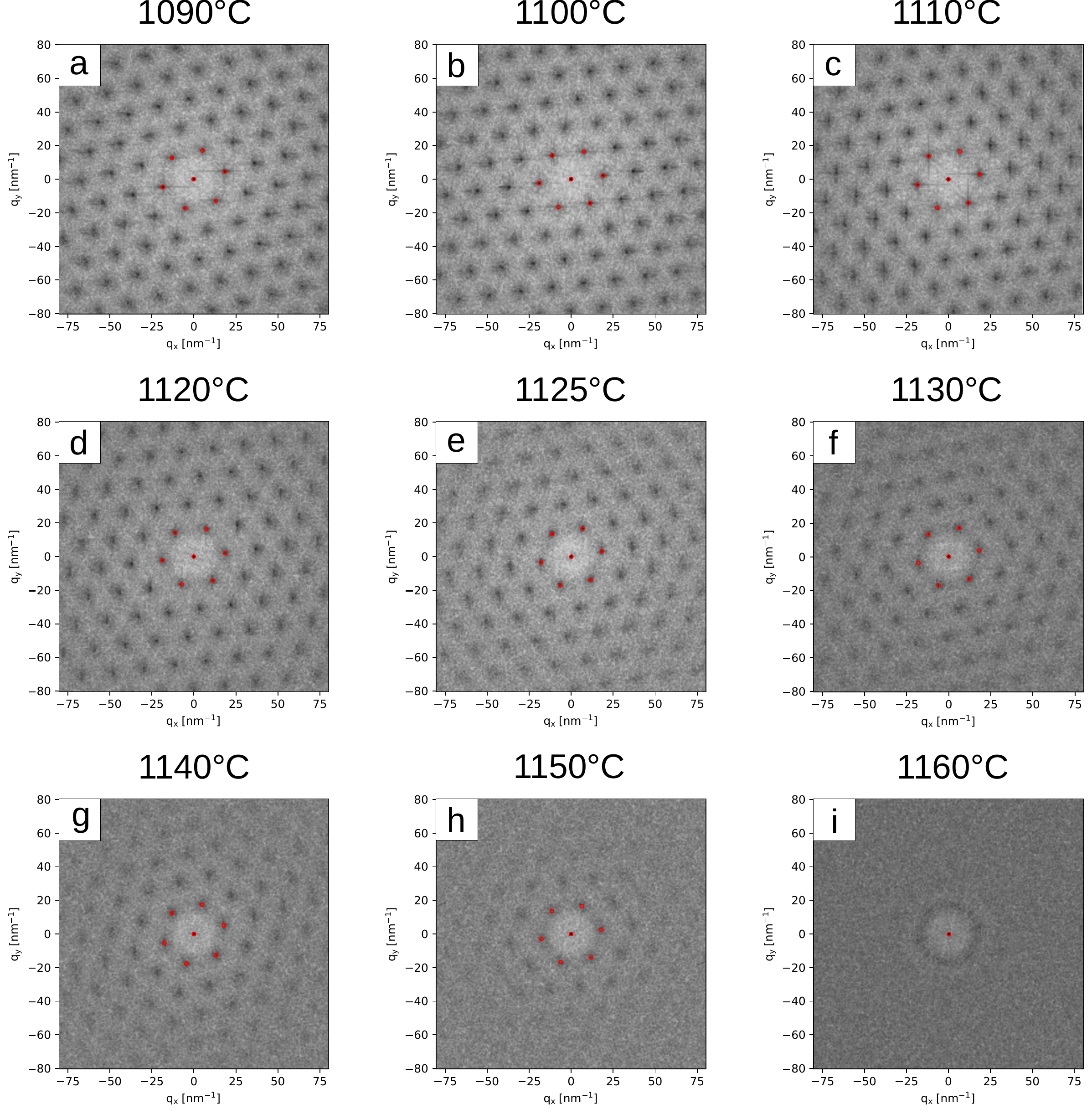}
    \caption{\textbf{2D structure factors S(q).} The 2D structure factors of individual frames at different temperatures calculated using the freud.diffraction module~\cite{ramasubramani_freud_2020}. The red circles mark the center positions of the 2D Gaussian fits. \textbf{a-e} Solid phase; \textbf{f-h} hexatic phase; \textbf{i-j} liquid phase.}
    \label{fig: SF}
\end{figure}
\newpage

\section*{On Possible Sources of Residual Error}

Vacancy-type defects with non-zero charge may induce slight in-plane distortions in AgI, potentially affecting the exponents of translational and rotational correlation decays \cite{ma_defect_2023}. However, as our baseline data shows, these defects are not activated in significant quantities by the electron beam, nor is there any indication they would dominate at higher temperatures. Despite this, the slow decay of the translational correlation at low temperatures reveals that the 2D AgI is not a perfect hexagonal crystal system. Contributing factors to this behavior likely include crystal anisotropy \cite{mustonen_toward_2022} and out-of-plane distortions \cite{hofer_picometer-precision_2023}, similar to those observed in 2D CuI. In some cases, these issues caused severe correlation decay even at room temperature, leading to the exclusion of certain crystals from our analysis early on. Another potential source of uncertainty is the ambiguity in the fitting process for the correlation functions, yielding parameter values $\eta_k$ and $\eta_6$ shown in Figure \ref{fig:Correlations}. However, the validity of the rotational correlation fit values is supported by the orientational parameter $|\Psi_6|$, which, based solely on hexagon positions and relative angles, exhibits behavior similar to $\eta_6$ (Figure~\ref{fig:Correlations}). Finally, inaccurate determination of reciprocal lattice vectors can lead to large deviations in $G_k$~\cite{li_accurate_2019}, and as such, all images without a clear crystal direction or identifiable 1st-order peaks were discarded in our analysis. This approach, however, likely leads to an underestimation of the translational decay exponent values at higher temperatures.
\clearpage
\section*{Orientational Parameters $|\Psi_6|$}
\begin{figure}[ht]
    \centering
    \includegraphics[width=12cm]{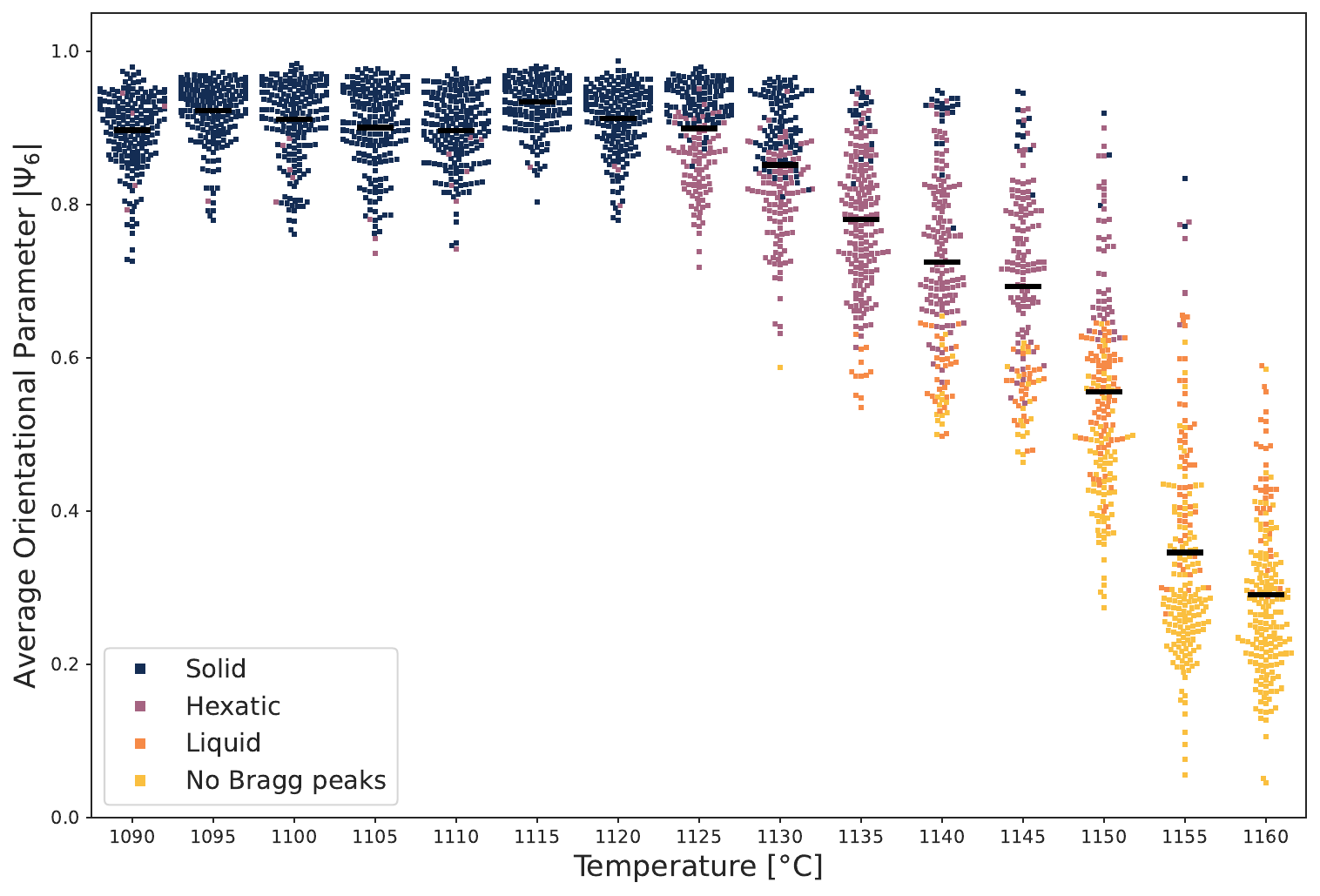}
    \caption{\textbf{Orientational parameters.} The average orientational parameter $|\Psi_6|$ as a function of temperature. The data points for the individual frames are color-coded according to the criteria described in Figure \ref{fig:Correlations}.}
    \label{fig: Orientational}
\end{figure}
\newpage
\section*{Supplementary Fourier Transforms}
\begin{figure}[ht]
    \centering
    \includegraphics[width=\textwidth]{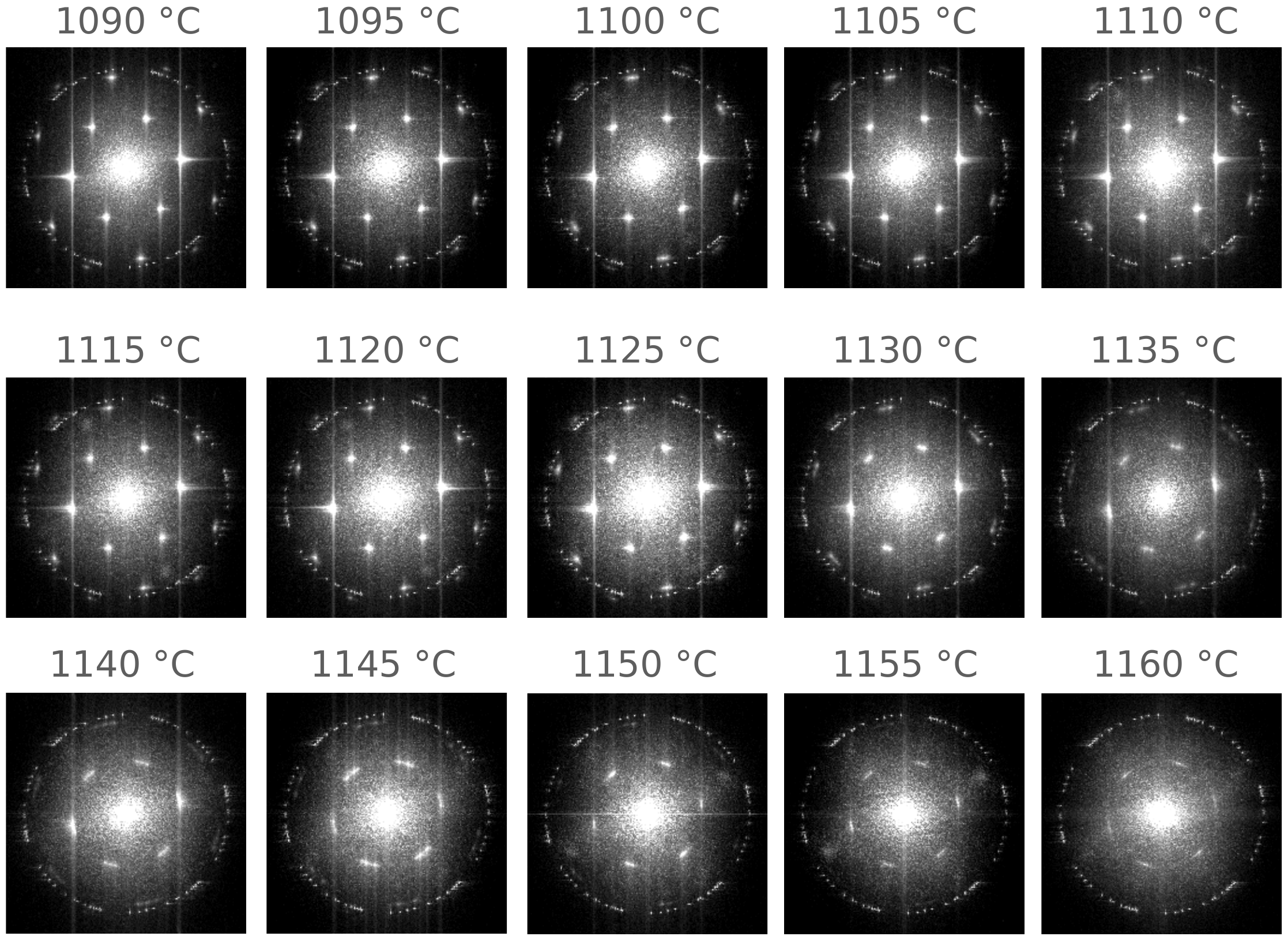}
    \caption{\textbf{Fourier transforms.} Fourier transforms (FT) of the data that were used to compute the spatial correlation functions shown in Figure \ref{fig:Correlations}. Each FT consists of an average of 200 STEM ADF images.}
    \label{fig:FT_Figure3}
\end{figure}
\newpage
\section*{ADF Image of the Crystal Studied for Manuscript Figures 3 and 4}
\begin{figure}[ht!]
    \centering
    \includegraphics[width=0.95\textwidth]{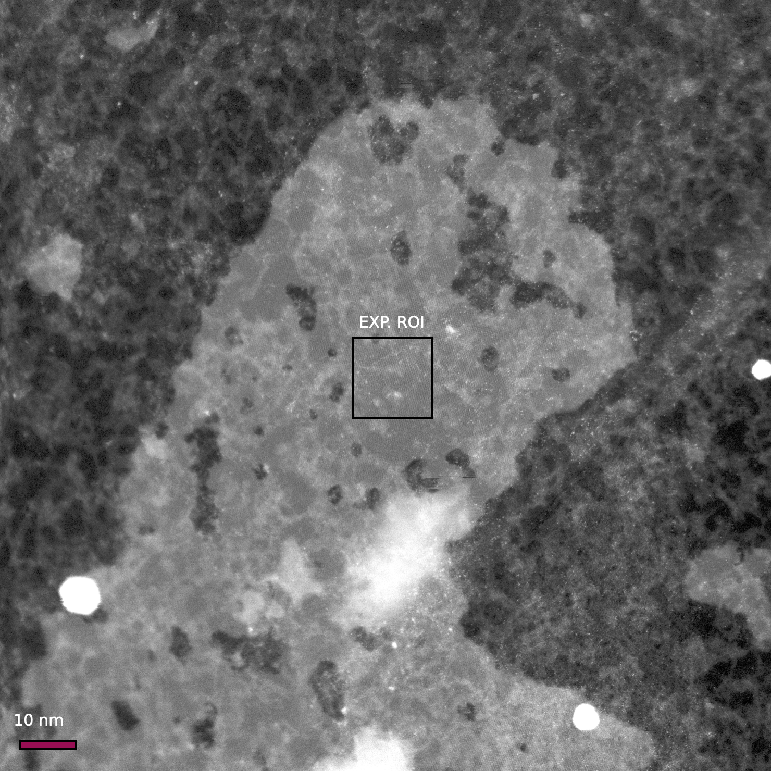}
    \caption{\textbf{An overview STEM-ADF image of the AgI crystal studied for Figures 3 and 4.} An overview image of the AgI crystal used to collect the data used to compute the spatial correlation functions and local density analysis in manuscript Figures \ref{fig:Correlations} and \ref{fig:density_distribution}. The ADF image was acquired after the melting experiment at room temperature. The data was collected from the area marked by a square box.}
    \label{fig:overview}
\end{figure}
\newpage


\end{document}